\documentclass[prd,singlecolumn,amsmath,nobibnotes,nofootinbib,superscriptaddress,preprintnumbers,floatfix]{revtex4}

\usepackage[usenames, dvipsnames]{color}
\usepackage{bm} 
\usepackage[linktocpage]{hyperref}
\usepackage{mathtools}
\RequirePackage{epsfig}
\usepackage{graphicx}
\usepackage{amsfonts}
\usepackage{physics}

\newcommand{\ie}{i.e.,\ }
\newcommand{\eg}{e.g.\ }

\newcommand{\ud}{{\rm d}}

\newcommand{\nodim}[1]{\ensuremath{\bar{#1}}}

\newcommand{\subPk}{\ensuremath{{\rm pk}}}
\newcommand{\subLR}{\ensuremath{{\rm L/R}}}
\newcommand{\subL}{\ensuremath{{\rm L}}}
\newcommand{\subR}{\ensuremath{{\rm R}}}
\newcommand{\subFv}{\ensuremath{{\rm fv}}}
\newcommand{\subnyq}{\ensuremath{{\rm nyq}}}
\newcommand{\subcut}{\ensuremath{{\rm cut}}}
\newcommand{\subTv}{\ensuremath{{\rm tv}}}
\newcommand{\peakRealization}{\ensuremath{\rho}_{\rm pk}}

\begin{document}

\title{Bubble Clustering in Cosmological First Order Phase Transitions}

\author{Dalila P\^irvu}
\email{dpirvu@perimeterinstitute.ca}
\affiliation{Perimeter Institute, 31 Caroline Street North, Waterloo, ON, N2L 2Y5, Canada}
\affiliation{Department of Physics and Astronomy, University of Waterloo,
Waterloo, Ontario, Canada, N2L 3G1}

\author{Jonathan Braden}
\email{jbraden@cita.utoronto.ca}
\affiliation{Canadian Institute for Theoretical Astrophysics, University of Toronto, Toronto, ON, M5S 3H8, Canada}

\author{Matthew C. Johnson}
\email{mjohnson@perimeterinstitute.ca}
\affiliation{Perimeter Institute, 31 Caroline Street North, Waterloo, ON, N2L 2Y5, Canada}
\affiliation{Department of Physics and Astronomy, York University, Toronto, ON, M3J 1P3, Canada}

\date{\today}

\begin{abstract}
False vacuum decay in quantum mechanical first order phase transitions is a phenomenon with wide implications in cosmology, and presents interesting theoretical challenges. In the standard approach, it is assumed that false vacuum decay proceeds  through the formation of bubbles that nucleate at random positions in spacetime and subsequently expand. In this paper we investigate the presence of correlations between bubble nucleation sites using a recently proposed semi-classical stochastic description of vacuum decay. This procedure samples vacuum fluctuations, which are then evolved using classical lattice simulations. We compute the two-point function for bubble nucleation sites from an ensemble of simulations, demonstrating that nucleation sites cluster in a way that is qualitatively similar to peaks in random Gaussian fields. We qualitatively assess the phenomenological implications of bubble clustering in early Universe phase transitions, which include features in the power spectrum of stochastic gravitational waves and an enhancement or suppression of the probability of observing bubble collisions in the eternal inflation scenario.
\end{abstract}

\maketitle

\section{Introduction}

Relativistic first order phase transitions take place in field theories with a potential that has a high energy false vacuum and a low energy true vacuum state; such theories are ubiquitous in physics beyond the Standard Model. Given a large region of space occupying the false vacuum, even at zero temperature, the false vacuum will decay via the formation of bubbles of the true vacuum~\cite{Kobzarev:1974cp,Coleman1977}. This is an important phenomenon with a wide range of cosmic implications. In the context of inflation, a postulated epoch of accelerated expansion in the early Universe, false vacuum decay was the original mechanism by which inflation was proposed to end~\cite{PhysRevD.23.347}. However, it was quickly realized~\cite{Guth1981a} that the phase transition from an inflating false vacuum to a non-inflating true vacuum does not complete unless the epoch of inflation is exceedingly short. Instead, inflation becomes eternal, only ending locally inside of isolated clusters of true vacuum bubbles (for a review, see e.g. \cite{Aguirre2007,Guth2007}). In the context of the string theory landscape, our observable Universe is thought to reside inside one such bubble.  The rare collisions between bubbles could provide observable evidence for this scenario~\cite{PhysRevD.76.123512,PhysRevD.76.063509,Chang:2007eq}. In the late Universe (e.g. after inflation), first order phase transitions can occur in models of baryogenesis~\cite{Kuzmin:1985mm,Cheung_2012}, string theory~\cite{Garcia_Garcia_2018}, supersymmetry~\cite{Pietroni:1992in,craig2009gravitational}, and even dark energy~\cite{Goldberg_2000,Dutta_2009}. A subset of these phase transitions can produce stochastic gravitational waves observable by gravitational wave detectors; see \eg~\cite{Christensen_2018} for a review.  

Vacuum decay is most frequently described in the literature using Euclidean instanton methods first developed by Coleman~\cite{Coleman1977, Callan1977}, which treat this problem in analogy with quantum tunnelling in non-relativistic quantum mechanics. Crucially, this framework cannot be applied to configurations with more than one bubble. Therefore, while the instanton methods can be used to compute decay rates (\ie the probability per unit 4-volume that a bubble will nucleate), one must assume that bubble nucleation events are independent in order to build a spacetime picture of the percolation of true vacuum bubbles. Testing this assumption is the primary goal of this paper. As we describe in more detail below, this assumption does not hold, but rather nucleation events are clustered in analogy with the clustering of rare peaks in Gaussian random fields.

Recently, a real-time semi-classical approach to vacuum decay was developed in~\cite{new_semiclassical_picture}. In this approach, the dynamical phase space evolution of a quantum state initially in the false vacuum is modelled using the truncated Wigner approximation \cite{PhysRev.40.749, gardiner_zoller_2004}. One generates an ensemble of initial conditions of the field and its conjugate momentum, drawn from the ground state defined by the false vacuum minimum. These initial states are then evolved classically with the non-linear Hamiltonian using lattice simulations. The field configuration in each realization is sampled at late times; in some realizations, bubbles of the true vacuum form. This procedure yields a semi-classical approximation to the first order phase transition dynamics leading to the decay of the false vacuum.  Although both the real-time semiclassical and instanton methods are semiclassical approximations it is still an open question precisely how these two approaches are related; see Refs.~\cite{new_semiclassical_picture,Hertzberg2019,Hertzberg:2020tqa,Blanco-Pillado:2019xny} for some discussion.  

In the limit where the nucleation rate is relatively slow (\eg compared to the light crossing time of the lattice), in any given realization the field behaves like a free scalar (with mass determined by the curvature about the false vacuum minimum) for much of its evolution. Starting from vacuum initial conditions, if the field truly were just a free massive scalar, then on each timeslice we obtain a Gaussian random field. It is well known that a Gaussian random field can be characterized by the properties of its maxima (or other extrema)~\cite{Bardeen1986,1984ApJ284L9K}. Peaks in a Gaussian random field cluster, and the properties of this clustering encode the power spectrum of the underlying field. This is the basis of cosmological studies using galaxies as a tracer of the underlying large-scale distribution of dark matter. Returning to our simulations, over a decorrelation timescale of order the inverse mass, the configuration of the field randomizes yielding a new set of peaks. Eventually, a peak will be high enough to leave the basin of attraction of the false vacuum, and the non-linearities of the potential allow something interesting to happen: the formation of a bubble inside of which the field settles into the true vacuum. If this bubble is large enough, it will expand. Within this picture, bubbles nucleate from peaks in the vacuum fluctuations of the field, and since peaks are clustered, bubble nucleation events should be clustered as well. Put simply, it is easier to nucleate bubbles from a region of space where the field is closer to the true vacuum. This runs contrary to existing work on relativistic first order phase transitions, which have implicitly assumed a distribution of bubbles statistically independent of position and time. 

In this work, we look more closely at the field region where the bubbles form and demonstrate that peaks of a critical spatial size and amplitude can act as seeds for bubble nucleation.  Using the nucleation seed properties to identify bubble sites, we compute the two-point correlation function between nucleation sites over an ensemble of 1+1-dimensional lattice simulations. We compare the result to the spatial correlation function for peaks in a scalar field with mass set by the false vacuum curvature, finding qualitative agreement, and validating the description above. 

There are a number of implications of a non-trivial two-point bubble correlation function:
\begin{itemize}
\item Just like galaxies provide a biased tracer of the underlying density field, bubble nucleation events provide a biased tracer of the underlying vacuum fluctuations. Just as a galaxy survey can be used to determine the statistics of the density field, the distribution of bubble nucleation events can be used to determine the statistics of vacuum fluctuations. The bubble correlation function is sensitive to deviations from the vacuum state, and would yield a different result for e.g. a thermal or vacuum state. There are associated connections with quantum measurement which deserve further exploration.
\item A number of analogue experimental systems that can be used as quantum simulators of false vacuum decay have been proposed~\cite{Fialko2015a,Fialko2017,Milsted:2020jmf}. The two-point bubble correlation function (or more generally, an n-point function) is an observable for these experiments, which could be used to confirm the validity of the semi-classical picture of vacuum decay and the properties of the initial state (as outlined in the point above).
\item  The frequency of collision between bubbles during eternal inflation and in post-inflation phase transitions is affected by the clustering of bubble nucleation sites. This has implications for programs to detect these effects in the CMB and spectrum of stochastic primordial gravitational waves.
\end{itemize}

The paper is organized as follows. In Section \ref{sec:VacuumDecay} we set up our field theory and explain how false vacuum decay arises from semi-classical evolution. In Section \ref{sec:PeakTh} we review Gaussian peak theory and discuss the general features of the two-point correlation function between peaks in one spatial dimension. In Section \ref{sec:BubbleTheory} we argue that bubbles form around peaks in the field and use empirical observations from our simulations to define bubble nucleation sites. In Section \ref{sec:Results} we present our results and compare with the free field theoretical prediction. In Section~\ref{sec:discussion} we discuss the implications of this result for bubble collisions and associated cosmological observables, and conclude in Section~\ref{sec:conclusion}.

\section{Real-time semiclassical formalism}\label{sec:VacuumDecay}

In this paper, we employ the real-time semiclassical simulation framework for scalar field theories that was shown to lead to vacuum decay in Ref.~\cite{new_semiclassical_picture}. We refer the reader to this paper for further details on the formalism, which we only briefly review here. Since our approach relies on running large ensembles of full nonlinear lattice simulations, we specialize to 1+1 dimensional systems for computational feasibility. However, we expect that the results presented below are qualitatively similar in higher spatial dimensions.

Consider a scalar field theory in 1+1 dimensions with Lagrangian density
\begin{equation}
    \mathcal{L} = \frac{1}{2} \dot{\phi}^2 - \frac{1}{2} \left( \partial_x \phi \right)^2 - V(\phi) \, .
\end{equation}
Note that in 1+1 dimensions, $\phi$ is dimensionless and $V(\phi)$ has mass dimension two. The fiducial potential $V(\phi)$ used in this paper is
\begin{equation}\label{potential_expr}
    V(\phi) = V_{0} \left[ -\cos \left( \frac{\phi}{\phi_0} \right)+\frac{\lambda^{2}}{2} \sin ^{2} \left( \frac{\phi}{\phi_0}\right) \right] \, .
\end{equation}
The parameter $\lambda$ modulates the depth of the false-vacuum potential well such that if $\lambda > 1$, the potential has an infinite periodic sequence of false minima at $\phi_{\subFv} = (2n+1)\pi \phi_0$, alternating with true minima at $\phi_{\subTv} = 2 n \pi \phi_0$, $n \in \mathbb{Z}$. In this paper, we focus on the regime where the decay rate of the false vacuum is relatively fast ($\lambda \sim 1$), so that we find a sufficient number of nucleation events  to empirically compute correlation functions.
The parameter $\phi_0$ scales the width of the potential, while $V_0$ scales its height.  $V_0/\phi_0^2$ sets the typical mass squared scales in the potential.  To adjust the amplitude of quantum effects, it is therefore convenient to vary $\phi_0$ while holding $V_0\phi_0^{-2}$ fixed.  This adjusts the width of the false vacuum minima relative to the typical amplitude of quantum fluctuations, which can be alternatively be viewed as adjusting $\hbar \propto \phi_0^{-2}$.
This form of the potential is motivated by proposals to simulate vacuum decay with cold atom experiments~\cite{Fialko_2017, Braden2018, Braden2018a,Billam:2020xna,Billam:2021qwt,Billam:2018pvp}. Our results are relevant to this program, but also generalize to any potential with multiple vacuum states. 

We initialize an ensemble of simulations with the mean field in the false vacuum $\phi = \pi\phi_0$, with fluctuations consistent with the ground state of the quadratic approximation to the false vacuum potential minimum 
\begin{equation}\label{fluctuations_gaussian}
    \phi(x, t=0) = \phi_{\subFv} + \delta \phi(x), \;\;\; \Pi(x, t=0) = \dot{\phi}(x,t=0) = \delta \Pi (x).
\end{equation}
The fluctuations $\delta \phi(x)$ and $\delta \Pi(x)$ are drawn from the Wigner functional of the initial state, with each initial draw corresponding to a single member of the ensemble.
Since we are interested in evolution from a false vacuum, we will approximate the initial Wigner functional as the vacuum associated with the quadratic expansion of the potential about the false vacuum. Investigating departures from this choice for the initial state is beyond the scope of this paper, but will be considered in future work. 
The initial fluctuations $\delta \phi$ and $\delta \Pi$ are drawn as realizations of Gaussian random fields whose Fourier coefficients $\delta\tilde{\phi}_{k}$ and $\delta\tilde{\Pi}_{k}$ have covariance
\begin{equation}\label{eq:powerspectrum}
    \ev{ \delta\tilde{\phi}_k^\ast \delta\tilde{\phi}_{k'} } = \frac{1}{2 \omega_k} \delta(k - k'), \ \ \     \ev{ \delta\tilde{\Pi}_k^\ast \delta\tilde{\Pi}_{k'} } = \frac{\omega_k}{2} \delta(k - k'), \ \ \ \ev{\delta\tilde{\phi}_k \delta\tilde{\Pi}_{k'}^*}=0 \, ,
\end{equation}
where $\omega_k^2=k^2 + m^2$ and $m^2 = m_{eff}^2 = V''(\phi=\pi \phi_0) = V_0 \phi_0^{-2} (\lambda^2-1)$.
Here $\left\langle\cdot\right\rangle$ represents an ensemble average, and we have assumed unitary normalization for the Fourier transforms in the continuum.

Each realization is evolved using the classical Hamilton's equations:
\begin{equation}\label{eomHamiltion}
    \frac{\ud \phi}{\ud t} = \Pi, \;\;\; \frac{\ud \Pi}{\ud t} =  \nabla^2 \phi - V'(\phi) \, .
\end{equation}
At a later time, we make measurements on the evolved ensemble. This approach captures the dynamical evolution of the Wigner functional~\cite{moyal_1949} to leading (nonperturbative) order in $\hbar$ of the form $e^{i / \hbar}$. Meanwhile, the leading perturbative quantum corrections are encoded in statistics of the initial fluctuations~\cite{new_semiclassical_picture} (see also \cite{Polkovnikov_2010,PhysRevD.50.7542}).
In the cold atom community, this is known as the Truncated Wigner Approximation (see e.g. Ref.~\cite{Blakie__2008} for a review), while in cosmology it is known as the stochastic lattice approximation and is used extensively in preheating studies (see \eg Ref.~\cite{Frolov:2010sz} for a review). For a free massive scalar, this procedure exactly describes the full quantum evolution for any initial state with positive definite Wigner functional, such as the vacuum state.  Including non-linearities in the potential, the story is more complicated. The classical non-linear time-evolution should capture all tree-level interactions between the Fourier modes. One complication arises from our need to initialize modes up to some cutoff $k_{\subcut}$ as introduced below---the effective dynamics of the longest wavelength modes on the lattice are modified from what we would expect in the bare potential alone (\ie renormalization effects will arise).
Nevertheless, in scenarios with unentangled initial states where quantum dynamics give rise to many particle, effectively classical final states, we can expect this procedure faithfully tracks the dynamics. One such situation is the decay of the false vacuum, which will be our main focus.

In order to implement the above procedure numerically, we must work with dimensionless spacetime coordinates and field variables.
As well, we must adapt the continuum prescription above to a discrete lattice of finite side length.
It is convenient to introduce dimensionless variables
\begin{equation}
    \nodim{t}=\mu t \quad \nodim{x}=\mu x \quad \nodim{\phi}=\frac{\phi}{\phi_0} \, ,
\end{equation}
where $\mu$ is some inverse length scale, and $\phi_0$ is as defined in the potential.
We assume throughout that $\hbar=c=1$, so that $\mu$ has units of mass and $\phi_0$ has units of $({\rm mass})^{(d-1)/2}$, where $d$ is the number of spatial dimensions.
In these units, the dimensionless equations of motion are
\begin{align}
  \frac{\ud\nodim{\phi}}{\ud\nodim{t}} &= \nodim{\Pi} \\
  \frac{\ud\nodim{\Pi}}{\ud\nodim{t}} &= \nodim{\nabla}^2\nodim{\phi} - \frac{V_0}{\mu^2\phi_0^2}\left[\sin\left(\bar{\phi}\right) + \frac{\lambda^2}{2}\sin\left(2\bar{\phi}\right)\right] \, .
\end{align}
We initialize the fluctuations on our finite discrete lattice of side length $L$ as
\begin{equation}
  \delta\nodim{\phi} \equiv \frac{\delta\phi}{\phi_0} =  \frac{1}{\phi_0\sqrt{\mu L}}\sum_{j=1}^{n_{\subcut}}\left[\frac{\hat{\alpha}_j}{\sqrt{2}}\sqrt{\frac{\mu^2}{V''(\phi_{\subFv})+k_j^2}} e^{ik_jx} + {\rm c.c.} \right] \, ,
\end{equation}
where $\hat{\alpha}_j$ is a realization of complex random deviate with variance $\left\langle\left|\hat{\alpha}_j\right|^2\right\rangle = 1$ and $k_j = \frac{2\pi}{L}j$.  Here we have truncated the spectrum at wavenumber $k_{\subcut} = \frac{2\pi}{L}n_{\subcut}$.
The initial realization of the momentum fluctuations is generated analogously
\begin{equation}
  \delta\nodim{\Pi} \equiv \frac{1}{\mu}\frac{\delta\dot{\phi}}{\phi_0} = \frac{1}{\phi_0\sqrt{\mu L}}\sum_{j=1}^{n_{\subcut}}\left[\frac{\hat{\beta}_j}{\sqrt{2}}\sqrt{\frac{V''(\phi_{\subFv})+k_j^2}{\mu^2}}e^{ik_jx} + {\rm c.c.} \right] \, ,
\end{equation}
with $\hat{\beta}_j$ a realization of a unit variance complex random deviate that is uncorrelated with $\hat{\alpha}_j$
The temporal evolution is performed using a 10th order accurate Gauss-Legendre scheme~\cite{Butcher:1964,Braden:2014cra}.
Spatial derivatives are computed by forward Fourier transforming, multiplying by the appropriate power of $ik$, then inverse Fourier transforming.
We refer to this as a Fourier pseudospectral approximation, and as a consequence the simulations have periodic boundary conditions.
We verified that the total field energy in the simulations is conserved to near machine-precision levels.

In the next section, we describe the picture expected for a free massive scalar, where the semiclassical approach described above is exact and we can compare numerical and analytical approaches. We then move on to a numerical study with the potential~\eqref{potential_expr}, where vacuum decay can occur.

\section{Peak-peak correlation function for a massive scalar}\label{sec:PeakTh}

In this section we describe the expected peak-peak correlation function for vacuum fluctuations in a free massive scalar field, which provides a warmup and a point of comparison for the analysis of the bubble correlation function. As described in the previous section, the initial condition is a spatially homogeneous Gaussian random field with spectrum given by Eq.~\eqref{eq:powerspectrum}. Evolving over a time of order $m^{-1}$, we obtain an uncorrelated Gaussian random field with the same power spectrum.  Therefore, evolving an ensemble of simulations with vacuum initial conditions simply propagates the vacuum. In addition to the two-point statistics (i.e.,\ power spectrum) of the field, we can compute the statistics of extremal points on a fixed time-slice. For the free field example, we focus on maxima (\ie peaks). The statistics of peaks in a Gaussian random field are described in the classic paper~\cite{Bardeen1986}. In this section, we review the derivation of the peak number density and of the two-point peak correlation function in one dimension. A similar derivation can be found in~\cite{Lumsden1989}. We then validate our numerical code by comparing with this expectation and show good agreement.

\subsection{Analytic derivation of the peak-peak correlation function}

Full realizations of our approximate false vacuum initial state possess fluctuations on all possible spatial scales, although the use of a discrete lattice enforces a truncation of this spectrum above some wavenumber below the Nyquist frequency.
However, we are primarily interested in variations of the field over distances of order $m^{-1}$.
Therefore, given a fine-grained realization of a Gaussian random field $\phi_{\rm fg}(x)$ (e.g. a realization of the field described in the previous section), it is natural to smooth it with a Gaussian kernel $W$ of width $R_0$
\begin{equation}
  \phi (x) \equiv \int \ud x' \ W(x,x')\phi_{\rm fg} (x') = \int {\rm d}x' \ \frac{e^{-\frac{(x-x')^2}{2R_0^2}}}{\sqrt{2\pi}R_0}\phi_{\rm fg}(x') \, .
\end{equation}
For the case of exploring peaks in the field, $R_0$ roughly corresponds to the width of peaks we are interested in.
Although many filters are possible, we choose a Gaussian filter because of its conceptual simplicity both as a local Gaussian smoothing in real space and a Gaussian truncation of high frequency modes in Fourier space.
We do not expect our qualitative conclusions to depend on this choice.
In a given field realization, the peaks in the smoothed field are distributed as a random process throughout space. A first guess may be that the peak locations are independent of each other. However, this is not quite correct, and there are important correlations between the peak locations as we review below.

We model the number density of peaks in a single field realization as a sum of Dirac delta functions
\begin{equation}
	\peakRealization(x) = \sum_{i} \delta(x-x_{\subPk,i}) \, ,
\end{equation}
where the $x_{\subPk,i}$ are the locations of peaks, labelled by the index $i$.
We want to relate the statistics of $\peakRealization$ (and $x_{\rm pk}$) to the underlying field $\phi(x)$.
 We denote the field, its gradient, and its curvature by $\phi(x)$, $\eta(x) = \phi^\prime(x)$, and $\zeta(x)=\phi^{\prime\prime}(x)$, respectively.  In the vicinity of a peak, we have $\phi(x) \approx \phi_{\subPk} +\frac{1}{2} \zeta_{\subPk}(x-x_{\subPk})^2$ and $\eta(x) \approx \zeta_{\subPk}(x-x_{\subPk})$.  Here we have indicated quantities evaluated at the location of the peak by ${\cdot}_{\subPk}$. It follows that $\delta(x-x_{\subPk}) = | \zeta(x_{\subPk}) | \delta(\eta(x))$.  The number density of maxima of $\phi$ where $\eta(x)=0$ and $\zeta(x)<0$ becomes $\peakRealization(x) = \left|\zeta(x)\right|\delta(\eta(x))$.  Therefore, to understand the statistics of individual peaks, it is convenient to first reduce the infinite dimensional space of field configurations down to the three-dimensional space of ${\bf y} \equiv (\phi,\eta,\zeta) = (\phi(x),\eta(x),\zeta(x))$ evaluated at a single point.  By translation invariance, the statistics of the random vector ${\bf y}$ over the ensemble of field configurations is independent of the choice of position $x$.  Peaks are selected by imposing appropriate constraints on $\eta$ and $\zeta$.  Similarly, peak-peak statistics can be tackled by considering the six dimensional random vector ${\bf y}_{2} = (\phi(x),\eta(x),\zeta(x),\phi(x+r),\eta(x+r),\zeta(x+r))$, which depends only on the separation $r$.

The statistics of ${\bf y}$ and ${\bf y}_2$ are specified by various two-point correlation functions between the field and its derivatives.  For future convenience, we therefore introduce
\begin{equation}\label{eq:moments}
    \sigma_{(m+n)/2}^2(r) = \ev{\partial^{(m)} \phi(x) \; \partial^{(n)} \phi(x+r)} = \begin{cases}
    \int\limits_{-\infty}^{+\infty} \dd{k}  k^{m+n} P(k) \cos{(k r)}, \text{ if } m+n \text{ even}\\
    \int\limits_{-\infty}^{+\infty} \dd{k}  k^{m+n} P(k) \sin{(k r)}, \text{ otherwise} \, .
    \end{cases}
\end{equation}
The case $r=0$ is sufficient to specify the distribution of single peaks, while the $r\neq 0$ information is required to explore peak-peak correlations.
Here $P(k)$ is the power spectrum of the smoothed field
\begin{equation}\label{eq:PSandFilter}
    P(k) = \left|W(k; R_0)\right|^2 P_{0}(k), \;\;\;\;\;W(k;R_0) = e^{-(k R_0)^2/2} \, ,
\end{equation}
where $P_{0}(k)$ is the power spectrum of the unsmoothed field and $W(k; R_0)$ is the Fourier transform of our Gaussian kernal with size $R_0$.
We will use the notation $\sigma_{(m+n)/2}^2(r)$ when $r\neq 0$, and $\sigma_{(m+n)/2}^2$ when $r=0$.

A first quantity of interest is the ensemble average peak density with height above a given threshold $\phi_{t}$
\begin{equation}\label{pknumberdensity}
	n_{\rm pk}(\phi_{t}) \equiv \ev{\peakRealization(x)} = \ev{\left|\zeta(x)\right|\delta(\eta(x)) } = \int_{\phi>\phi_{t}} \int_{\zeta<0} \mathcal{P}(\phi,\eta,\zeta;\mathcal{M}) \Big|_{\eta=0}  \left|\zeta\right| \ud\zeta \ud\phi,
\end{equation}
where $\mathcal{P}(\phi,\eta,\zeta;\mathcal{M}) \ud\phi \ud\eta \ud\zeta$ is the joint probability distribution for the random variables $\vec{y} = (\phi,\eta,\zeta)$ evaluated at a single spatial position.
The conditions $\eta=0$ and $\zeta <0$ select maxima of the field, but the statistics of maxima are no different than that of the minima. Finally, the number density depends on the peak amplitude threshold $\phi_t$.
Since we are ultimately interested in ``peaks'' probing the nonlinear structure of our false vacuum potential~\eqref{potential_expr}, it is convenient to consider the rescaled threshold $\nodim{\phi}_t = \phi_t/\phi_0$.

All that remains is to obtain the probability density $\mathcal{P}$.  It is straightforward to see that $(\phi,\eta,\zeta)$ are jointly Gaussian distributed
\begin{equation}\label{gaussianprobabilityfunction}
	\mathcal{P}(\phi,\eta,\zeta;\mathcal{M}) \ud\phi \ud\eta \ud\zeta = \frac{e^{-\frac{1}{2} \mathbf{y}^{\mathrm{T}} \cdot \mathcal{M}^{-1} \cdot \mathbf{y}}}{\sqrt{(2\pi)^3 \det \mathcal{M}}} \ud\phi \ud\eta \ud\zeta \, .
\end{equation}
The probability density is specified by the covariance matrix $\mathcal{M}$ with elements $m_{ab} \equiv \ev{y_a y_b}$, where ${\bf y}=(\phi, \eta, \zeta)$ as above.
Using~\eqref{eq:moments}, we thus have
\begin{equation}
    \mathcal{M} = \begin{pmatrix}
    \sigma_0^2 & 0 & -\sigma_1^2 \\
    0 & \sigma_1^2 & 0\\
    -\sigma_1^2 & 0 & \sigma_2^2 
    \end{pmatrix}
\end{equation}
and~\eqref{pknumberdensity} becomes
\begin{equation}\label{peak_numdens}
  n_{\mathrm{pk}}\left( \nodim{\phi}_{t} \right) = -\frac{1}{\sqrt{(2\pi)^3\sigma_1^2\left(\sigma_0^2\sigma_2^2-\sigma_1^4\right)}}\int_{\nodim{\phi}_{t}\phi_0}^{+\infty} \int_{-\infty}^{0} e^{-\frac{1}{2}{\bf y}^T\cdot\mathcal{M}^{-1}\cdot{\bf y}|_{\eta = 0}} \zeta \ud\zeta \ud\phi \, ,
\end{equation}
with
\begin{equation}
  \mathbf{y}^{\mathrm{T}} \cdot \mathcal{M}^{-1} \cdot \mathbf{y} \Big|_{\eta=0} = \frac{\sigma_2^2 \phi^2 + 2 \sigma_1^2 \phi \zeta + \sigma_0^2 \zeta^2}{\sigma_0^2 \sigma_2^2 - \sigma_1^4} \, .
\end{equation}
In the limit of high $\nodim{\phi}_{t} \gg \sigma_0/\phi_0$, the dominant term in the exponent is $-\phi^2/\sigma_0^2$. We therefore see that the number density of peaks decreases as the threshold is increased.

Now we turn to the derivation of the peak-peak correlation function. It describes the clustering of peaks, which are a biased tracer of the underlying field. Excluding self-pairs, the two-point correlation function $\xi_{\mathrm{pk}}$ is defined in terms of the peak number density as
\begin{equation}\label{xi_definition}
   n_{\subPk}^2(\nodim{\phi}_{t}) \left( 1 + \xi_{\subPk}(r) \right) = \ev{\peakRealization(x)\peakRealization(x+r)},
\end{equation}
which is the joint probability that peaks exist in two volume elements separated by a distance $r$ divided by the square of the overall peak number density at fixed threshold $\nodim{\phi}_{t}$.
Denoting the properties of the peak at $x$ and $x+r$ with the subscript ${}_1$ and ${}_2$, a point in configuration space is now specified by the six dimensional vector ${\bf y}_2 = \left( \phi_1,\eta_1,\zeta_1,\phi_2,\eta_2,\zeta_2 \right)$.
As in the case of the single peak parameter space, these variables are jointly Gaussian distributed.
To investigate the correlation between pairs of maxima that both exceed the same threshold $\phi_t$, we enforce $\left\{ \phi_1>\phi_t, \eta_1=0, \zeta_1<0, \phi_2>\phi_t, \eta_2=0, \zeta_2<0 \right\}$.
Equation \eqref{xi_definition} can be rewritten as
\begin{equation}\label{full_pkpk_corr}
    1+\xi_{\subPk}(r)= \frac{n_{\subPk}(\nodim{\phi}_t)^{-2} }{ (2\pi)^3 \sqrt{\det \mathcal{M}_{\rm pair}}} \iint_{\nodim{\phi}_t \phi_0}^{\infty} \iint_{-\infty}^{0} e^{-\frac{1}{2} \mathbf{y}_2^{\mathrm{T}} \cdot \mathcal{M}_{\rm pair}^{-1} \cdot \mathbf{y}_2 }  \zeta_1 \, \zeta_2 \, \ud\zeta_1 \ud\zeta_2 \ud\phi_1 \ud\phi_2.
\end{equation}
Here, the correlation matrix for the pair of peaks is given by
\begin{gather}
    \mathcal{M}_{\rm pair}(r) = \begin{pmatrix}
    \mathcal{M}_{11} & \mathcal{M}_{12}(r) \\
    \mathcal{M}_{21}(r) & \mathcal{M}_{22}
    \end{pmatrix},\\
    \mathcal{M}_{11} = \mathcal{M}_{22} = \begin{pmatrix}
    \sigma_0^2 & 0 & -\sigma_1^2 \\
    0 & \sigma_1^2 & 0 \\
    -\sigma_1^2 & 0 & \sigma_2^2 
    \end{pmatrix},\,\,\,\,\,
    \mathcal{M}_{12}(r) = \mathcal{M}_{21}^*(r) = \begin{pmatrix}
    \sigma_0(r)^2 & -\sigma_{1/2}(r)^2 & -\sigma_1(r)^2 \\
    \sigma_{1/2}(r)^2 & \sigma_1(r)^2 & -\sigma_{3/2}(r)^2 \\
    -\sigma_1(r)^2 & \sigma_{3/2}(r)^2 & \sigma_2(r)^2 
    \end{pmatrix} \, .
\end{gather}
From this we obtain the exponent in~\eqref{full_pkpk_corr} as
\begin{align}
  \mathbf{y}_2^{\mathrm{T}} \cdot \mathcal{M}^{-1} \cdot \mathbf{y}_2 = m_{11}^{-1}& \left(\phi_1^2+\phi_2^2\right)+2 m_{14}^{-1} \phi_1\phi_2+m_{33}^{-1} \left(\zeta_1^2+\zeta_2^2\right)+ 2 m_{36}^{-1} \zeta_1 \zeta_2 \notag \\
  &- 2 m_{13}^{-1}\left(\phi_1\zeta_1+\phi_2\zeta_2\right)-2 m_{16}^{-1} \left(\phi_1 \zeta_2 +\phi_2 \zeta_1\right) \, ,
\end{align}
where $m_{ab}^{-1}$ be elements of the $6\times6$ matrix $\mathcal{M}_{\rm pair}^{-1}$. The matrix $\mathcal{M}_{\rm pair}^{-1}$ is singular at $r=0$, but well defined for $r$ positive.  For the scales below the smoothing scale $R_0$, $\xi_{\subPk}(r \to 0) = -1$. At small separations the dominant term in the exponent has prefactor $m_{11}^{-1} \sim r^{-2}$, so $\xi_{\subPk}(r)$ picks up as $e^{-c/r^2}$ for some constant $c$.  For large separations the exponential inside the integral dominates and $\xi_{\subPk}(r \gg R_0)$ plateaus around a constant value. In the limit of large $\nodim{\phi}_{t} \gg \sigma_0/\phi_0$, the main contribution to $\xi_{\subPk}$ at intermediate separations is from $n_{\subPk}(\nodim{\phi}_t)^{-2}$ which grows as $e^{2 \nodim{\phi}_t^2 \phi_0^2}$.
Therefore, peaks of greater height (\ie increasingly rare) cluster more strongly than peaks of lower height. Examining the falloff of $\xi_{\subPk}(r)$ in greater detail, the correlation length is of order a few times $m R_0$.

When we compare the prediction for the peak-peak correlation function of the formalism in the continuum limit with the results from simulations, we must use the discrete version of the field power spectrum. This is so that we obtain an analytic result that matches the implementation on the discrete lattice of the field. In practice this means replacing the integrals in \eqref{eq:moments} with a sum over all modes
\begin{equation}
	\sigma_{(m+n)/2}^2(r_{ij}) = \begin{cases}
		\sum\limits_{\substack{l=-n_{\subcut}\\l\neq0}}^{n_{\subcut}} k_l^{m+n} P(k_l) \cos{(k_l r_{ij})}, \text{ if } m+n \text{ even}\\
		\sum\limits_{\substack{l=-n_{\subcut}\\l\neq0}}^{n_{\subcut}} k_l^{m+n} P(k_l)  \sin{(k_l r_{ij})}, \text{ otherwise} \, ,
	\end{cases}
\end{equation}
where we assume either that the spectral cut is below the Nyquist frequency, or else that we have an odd number of grid sites, so that the lower limit of the sum extends to $-n_{\rm cut}$.

\subsection{Numerical peak-peak correlation function}
We now wish to demonstrate that we can reproduce the peak-peak correlation function empirically using an ensemble of our lattice simulations. This is an important step to confirm that our simulations accurately time-evolve the peak statistics, in preparation for studying the bubble-bubble correlation function. 
Since our initial conditions are Gaussian, and Gaussianity is maintained by linear field evolution, in this section we consider a free massive scalar field with potential
\begin{equation}\label{potential_free}
    V(\phi) = \frac{1}{2} m^2 (\phi - \phi_{\subFv})^2 = \frac{1}{2}m^2\phi_0^2\left(\frac{\phi}{\phi_0} - \frac{\phi_{\rm fv}}{\phi_0}\right)^2 \, ,
\end{equation}
with initial conditions corresponding to Minkowski vacuum fluctuations with mass $m$.
For convenience, and ease of comparison with later results, we choose a mass consistent with the second derivative of the bare potential~\eqref{potential_expr} about the false vacuum, $m^2 = V_0\phi_0^{-2}(\lambda^2-1)$. The field and momentum are initialized as described in Section~\ref{sec:VacuumDecay}.
From~\eqref{full_pkpk_corr}, we see that the predicted peak-peak correlation is determined by the peak threshold $\phi_t$, as well as a few lower order sinusoid weighted ``moments'' of the field power spectrum.
For our initial conditions, the moments are in turn specified by the field mass $m$ and filter smoothing scale $R_0$ in the combination $mR_0$, and the field scale $\phi_0$.  A key quantity characterizing the rarity of the peaks is $\phi_t^2/\sigma_0^2 \propto \phi_0^2$.

\begin{figure}
    \centering
    \includegraphics[width=0.6\textwidth]{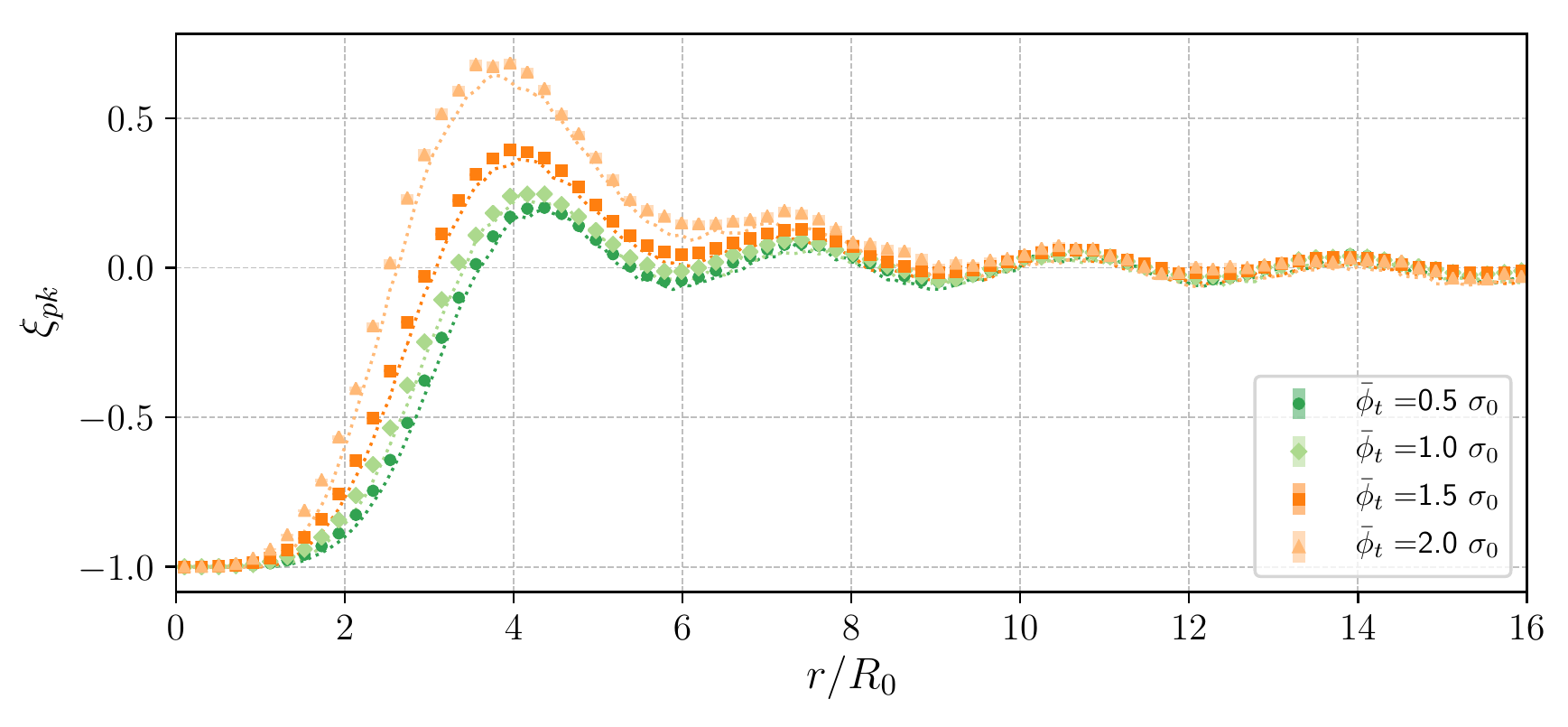}
    \caption{The equal-time peak-peak correlation function for four detection thresholds $\nodim{\phi}_t$ computed analytically (dashed lines) and extracted from an ensemble of simulations (points). The agreement between analytic and numerical results serves as a validation exercise for the simulations. The thresholds are chosen in units of the field RMS $\sigma_0$. Larger values correspond to a larger peak amplitude in $\xi_{\text{pk}}$ showing that the tallest peaks cluster more strongly. $\xi_{pk} = -1$ where the separation $r$ is less than the smoothing scale $R_0$. It picks up exponentially to reach a maximum at separation $r \sim 4 R_0$, the scale where peaks are the most frequent. This balances out at larger separations where $\xi_{pk} < 0$. Note that the lattice size is $r_{max}/R_0 \sim 415$. The small mis-match between the analytic curves and simulations arises due to a small bias in the estimator Eq.~\eqref{rivolo} from realizations with few peaks.}
    \label{fig:free_field_corr}
\end{figure}

Evolving the field, we obtain $\xi_{\subPk} (r)$ on a single randomly selected timeslice in each simulation by detecting all peaks above a fixed threshold $\nodim{\phi}_t$. We then compute the equal-time peak-peak correlation function using the estimator~\cite{Rivolo1986}:
\begin{equation}\label{rivolo}
	1 + \xi_{\subPk}(r)=\left\langle\frac{1}{K} \sum_{i=1}^{K} \frac{K_{i}(r)}{n_{pk} V_i} \right\rangle \, ,
\end{equation}
where $K_i (r)$ is the number of peaks inside a search volume $\Delta V_i = 2 \Delta r \phi_0^{-1}\sqrt{V_0}$ at spatial separation within $\pm [r,r+\Delta r) \phi_0^{-1}\sqrt{V_0}$ from the reference peak $i$. The factor of $2$ arises from the fact that we search for peaks both to the left and to the right of the reference peak $i$ up to separations $L/2$. $n_{\subPk}$ is the measured number density of peaks on the timeslice, and $K$ the total number of peaks on that slice, for a given realization. In other words, the correlation function is the expected ratio of the number density of peaks a distance $r$ from a randomly chosen peak to the mean number density. The $\ev{\cdot}$ denotes the ensemble average of the correlation function, which in this section we take to be $200,000$ simulations. The result for fixed mass and smoothing scale, but varying threshold, is shown in Fig.~\ref{fig:free_field_corr}, where the smoothing scale is $m R_0 \sim 1/3$, and $k_{\subcut} R_0 \sim 2$. Data is binned such that $\Delta r/R_0 \sim 0.2$. The lattice parameters are consistent with those described in detail in Section~\ref{sec:BubbleTheory} for the simulations including false vacuum decay. We compare the empirically measured correlation function to the analytic expectation from Eq.~\eqref{full_pkpk_corr}. The prediction and lattice simulation results agree up to a small sampling bias at higher thresholds arising due to a large sample variance on the realization peak number density. We find that the distribution is independent of timeslice, confirming our expectation that time evolution merely propagates the vacuum statistics.

A few properties of the correlation function are noteworthy. First, it can be seen that the peak amplitude of the correlation function increases with threshold, holding to the expectation that rarer peaks cluster more strongly. The increasing noise and bias on the data points for higher thresholds in Fig.~\ref{fig:free_field_corr} is due to increasing sample variance (\ie rarer peaks). A consistency check on the estimator Eq.~\eqref{rivolo} is to ensure that
\begin{equation}
\int_{-L/2}^{L/2} \dd{r} \ (1 + \xi_{pk}(r)) = 1 \, ,
\end{equation}
to good approximation. In other words, if peaks are correlated at short separation then to maintain the average number density throughout the volume they must be anti-correlated at large separation. We verify that this is indeed true for our ensemble of simulations, at all times.

Before moving on to discuss vacuum decay, let's consider the following thought experiment. Imagine that we had a physical system consisting of a scalar field in its vacuum state.  Now suppose we could construct a `peak detector' that was sensitive to peaks in the (spatially) smoothed vacuum fluctuations above some threshold $\nodim{\phi}_t$, enabling one to replicate our numerics with an experimental protocol. For example, one could imagine that the value of $\phi$ controls the lifetime of some unstable particle such that when $\nodim{\phi} > \nodim{\phi}_t$ decay happens very quickly, but is shut off when $\nodim{\phi} < \nodim{\phi}_t$. Filling space with a dilute gas of such particles (with the inverse number density roughly corresponding to the smoothing scale), we could then detect the position of peaks by locating the origin of the detected decay products of the unstable particles. From this distribution of peak positions, we could compute the two point correlation function using the estimator Eq.~\eqref{rivolo}. Finally, since the spatial dependence of the peak-peak correlation function is in one-to-one correspondence with the power spectrum of the vacuum fluctuations, it is possible to extract the statistics of the vacuum fluctuations underlying the peaks (at least in the infrared)~\footnote{The interaction with the unstable particle will alter the vacuum state of the massive scalar, so one must be a bit careful with this thought experiment to be precise about which state one is probing. Here, we assume it is the state defined by the non-interacting vacuum of the massive scalar.}. As we will see in the next section, false vacuum decay appears to be analogous to this example. 

\section{False vacuum decay}\label{sec:BubbleTheory}

We now move on to discuss vacuum decay in the potential Eq.~\eqref{potential_expr}. Our ultimate goal is to compute the two point correlation function between bubble nucleation sites as a function of both their spatial and temporal separations. The following two sections address this. In this section, we review the spacetime picture of false vacuum decay and construct an algorithm for identifying bubble nucleation sites. In the next section we will apply this formalism to bubble nucleations in our numerical simulations.

We initialize an ensemble of simulations as described in Sec.~\ref{sec:VacuumDecay}. The mean field is at the false vacuum minimum $\nodim{\phi}_{\subFv}=\pi$, and fluctuations about the false vacuum evolve non-linearly according to the equations of motion. Every once in a while, the non-linear evolution leads to large localized fluctuations in the field.  Some of these fluctuations overcome the potential barrier, and from here the field may either bounce back or continue to roll down the potential into one of the true vacua, depending on the size of the spatial region where the fluctuation occurs. In the latter case, a bubble of true vacuum is formed, which subsequently expands. We confirm below that this is a reasonably good proxy for bubble formation. Each realization experiences a different number of decay events, and the full ensemble is used to extract statistics about the bubble formation process. 

\begin{figure}
    \centering
    \includegraphics[width=0.4\textwidth]{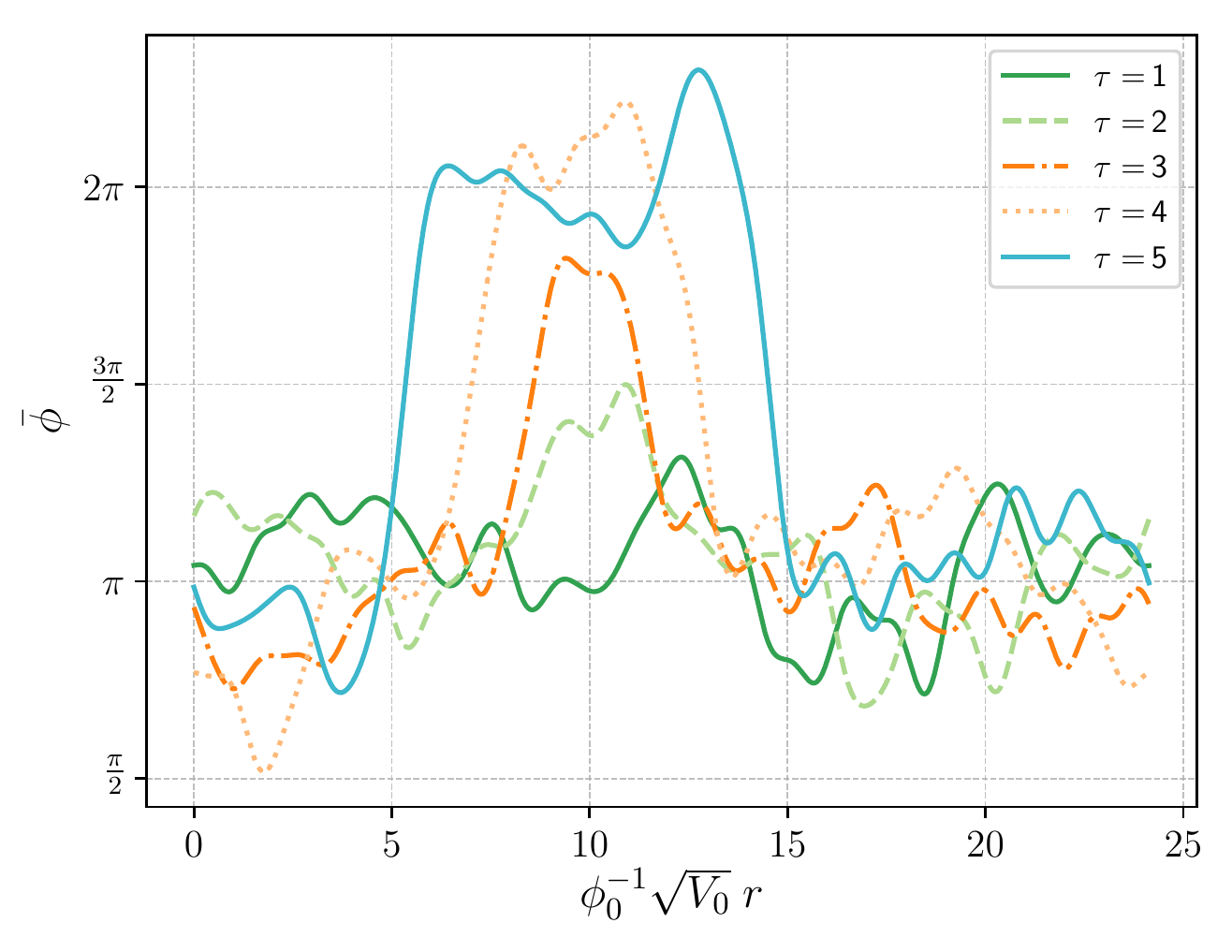}
    \caption{Stages in the evolution of a bubble, as snapshots of the smoothed field amplitude over a lattice region. The snapshots are spaced by equal-time intervals. One can follow time backwards and trace the bubble to the location of a peak rising gradually above the mean field.}
    \label{fig:bubbleevol}
\end{figure}

A series of timeslices around a typical nucleation event found in the simulations is shown in Fig.~\ref{fig:bubbleevol}. Not every peak triggers a nucleation event; it must be of sufficient amplitude and width to do so. This led us to consider extrema in a smoothed field as a proxy for nucleation events. Since there are two true vacua at $\nodim{\phi}_{\subTv} = 0 , \ 2 \pi$, maxima about the false vacuum decay to $\nodim{\phi}_{\subTv} = 2\pi$ while minima decay to $\nodim{\phi}_{\subTv} = 0$. 

In order to remove the small scale noise while still resolving the structure of individual bubbles, we want to define our smoothing scale to be slightly smaller than the typical size of a bubble early in its time evolution. We empirically determine the typical bubble size by stacking bubbles from different simulations in our ensemble to find the mean bubble profile.
An immediate challenge is that most bubbles are not formed at rest, and many have a center of mass velocity that is highly relativistic. The shape of the bubble walls and the bubble size in the frame of reference of the simulation is distorted by relativistic length contraction. To obtain the average bubble, we have to bring the bubbles to rest before averaging. The detailed properties of the average bubble and the distribution of velocities are of general interest for understanding the properties of vacuum decay. Here, we simply seek to motivate a smoothing scale, and defer a more comprehensive investigation to future work. Note that although we define the smoothing scale by the size of the average bubble in its rest frame, relativistic bubbles which are length contracted can still be detected. We elaborate upon this below.

We prepare an ensemble of simulations with $\lambda=1.5$ and $\phi_0 = 2.22$. We initialize the field fluctuations with this larger value of $\phi_0$ (and hence lower amplitude) in order to decrease noise around the bubble walls. In turn, this lowers the decay rate, allowing us to more easily isolate individual bubbles.  If we vary $\phi_0$ while holding $V_0/\phi_0^2$ fixed, the shape of the bubble is independent of $\phi_0$ at tree level, allowing us to use the same average bubble size for all the simulations presented below. We initialize the field evolution using the same grid parameters as the non-linear simulations, which are defined below. Nucleation events are identified in the simulations where the field has transitioned to the true vacuum, and bubbles are extracted. These bubbles materialize with a spread of velocities ranging between zero and nearly the speed of light. In order to bring a bubble into its rest frame, we extract the velocity of the walls (which follow hyperbolic trajectories) from which we obtain the centre of mass velocity of the bubble $v_{\rm COM}$, and apply an inverse Lorentz boost in order to bring the bubble into its rest frame. We start by obtaining the $r_{\subL}(t)$ and $r_{\subR}(t)$ trajectories for the centres of the left- and right-travelling bubble walls respectively, on each timeslice, over the entire extent of the bubble, and most importantly around the moment of nucleation where the vertex is located. We proceed as follows:
\begin{enumerate}
	\item We obtain $r_{\subLR}$ in each timeslice $t$ as the best fit parameters of the field value $\nodim{\phi}(t, r)$ to the expression $\pm \left( \tanh{ \frac{r-r_{\subL}}{w_{\subL}}} + \tanh{\frac{r_{\subR}-r}{w_{\subR}}}\right) \frac{\nodim{\phi}_{\subFv}}{2} + \nodim{\phi}_{\subFv}$, where $r$ is a coordinate that spans the lattice and $w_{\subLR}$ is a measure of the wall thickness. We impose that $w_{\subL} = w_{\subR}$ in the rest frame.
	\item With the values $r_{\subLR}$ obtained this way, we fit each wall individually to a hyperbola $r_{\subLR}(t) = \pm \sqrt{a_1 + (t - a_2)^2} + a_3$ with free parameters $a_1, a_2, a_3 \in \mathbb{R}$ to get the full trajectory. The wall velocity is $v_{\subLR} = \dot{r}_{\subLR}(t)$.
	\item Using relativistic velocity addition we obtain the centre of mass velocity for the entire bubble $v_{\rm COM}$, and the instantaneous wall velocity $v_{\rm wall}(t)$ in the bubble's rest frame. The Lorentz boost factor is the value of $v_{\rm wall}(t)$ that minimizes the difference $|v_{\rm COM} - v_{\rm wall}(t)|$.
	\item We apply a Lorentz boost transformation on the grid coordinates, deforming the bubble by interpolating it onto the new grid.
	\item This process is iterated until the $v_{\rm COM}$ of the transformed bubble is below $0.1c$, with $c$ the speed of light on the lattice. Due to computational limitations imposed by our interpolating scheme, the maximum boost factor on each iteration we apply is $v=0.9c$.
\end{enumerate}

\begin{figure}
	\centering
	\includegraphics[width=0.75\textwidth]{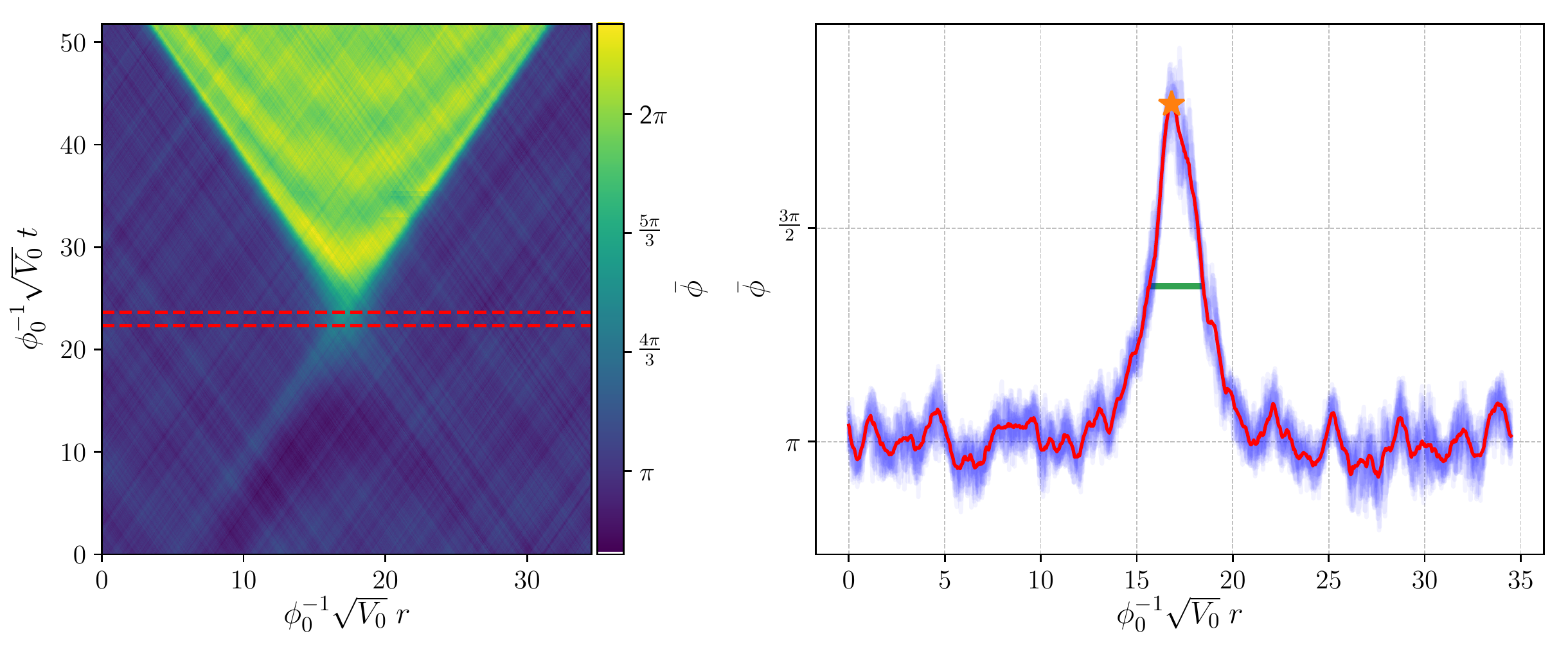}
	\caption{On the left we show the averaged bubble in its rest frame. We construct this by de-boosting 15 bubbles to their rest frame, translating the nucleation center to a specified location, and averaging. On the right we show in blue the field profile of the average bubble for all the timeslices highlighted between the red curves on the left. These correspond to the interval where the wall velocity is $\leq 25\%$ of the speed of light on the lattice. The red curve shows the time average of these profiles. The green line shows the FWHM of the red curve, relative to the maximum highlighted by the orange star. The average bubble radius $\nodim{R}$ is taken to be a half FWHM.} \label{fig:average_thick_bubble}
\end{figure}

Applying this procedure to our simulations, we obtained $15$ stationary bubbles. Ten of these were found to have nucleated with relativistic velocities $v>0.5c$, with $3$ of these having $v>0.9c$. The average bubble profile is shown in Fig.~\ref{fig:average_thick_bubble}. We identify the time interval where the instantaneous wall velocity is $\leq 25\%$ of the speed of light.  We extract the time-averaged field profile for the rest-frame average bubble over this interval. We then obtain the FWHM of this field profile and identify average bubble radius $\nodim{R}$ with one half of this value. This procedure is pictured in Fig.~\ref{fig:average_thick_bubble} Once a bubble materializes in simulations in its own rest frame with the above width and height, it will continue to grow and expand relativistically until it stabilizes in amplitude around the true vacuum. 

Having defined a smoothing scale to apply to the simulation data, we now define an algorithm for identifying the peaks in the field that correspond to nucleation sites for bubbles decaying to $\nodim{\phi}_{\subTv} = 2\pi$ as follows:
\begin{enumerate}

\item We smooth the field in each realization along the spatial axis on each timeslice. The smoothing is done with a Gaussian kernel of width $R_0$, as we do for the case of the free field (see \eqref{eq:PSandFilter}). We choose two smoothing scales for comparison: $R_0 = \nodim{R}/4$ and $R_0 = \nodim{R}/2$, where $\nodim{R}$ is the average radius of bubbles. Fluctuations over significantly smaller spatial regions do not seed bubble nucleation events. 

\item To eliminate bubbles nucleating directly from peaks in the initial conditions, we discard all events detected to occur for $t \leq 2 m^{-1}$. For a free field, this is of order the decay time for peak-peak correlations. The remaining events occur from nonlinear processing encoded in the equations of motion.

\item Stepping through each subsequent timeslice, we locate extrema in the smoothed field that exceed a threshold $\nodim{\phi} > \nodim{\phi}_t$, $\nodim{\phi}_t = \nodim{\phi}_{\rm max} + \Delta |\nodim{\phi}_{\rm max} - 2\pi |$, with $\nodim{\phi}_{\rm max}$ the value of the field at the potential maximum to the right of the false vacuum, and $\Delta \in \{0.2, 0.25, 0.3\}$. These values for $\Delta$ are chosen as they give a $\nodim{\phi}_t$ greater than the amplitude of the average bubble at nucleation. We compare three choices for threshold below, and confirm that our results are not highly sensitive to the value chosen over a significant range. 

\item To confirm that a peak is indeed a nucleation event, we check the field evolution within the future lightcone of the detection site.  In order for a detection site to be classified as a nucleation event, we require that the field within the future light cone both exceeds detection threshold over a period $\sim m^{-1}$ and remains within the true vacuum well. Specifically, if the field peak returns to the original false vacuum we consider that the bubble has dispersed. Alternatively, if the amplitude reaches a different false or true vacuum, we consider that the signal corresponds to a collision of the bubble wall with large fluctuations in the background field, or with another bubble. These cases are excluded by imposing amplitude cuts at $\nodim{\phi}_t \leq \nodim{\phi} \leq 2\pi$.

\item Once a nucleation event is located, we do not assign further nucleation events within its future lightcone or within the bounds of the bubble at subsequent times (the bubble associated to a nucleation event is defined as the field region where $\nodim{\phi} > \nodim{\phi}_t$). Imposing a maximal amplitude cutoff as described in the previous step is also useful way of discarding false signals that mimic a nucleation event around the bubble wall, which occur marginally outside of the future lightcone of the nucleation sites and would otherwise add noise to our data.
\end{enumerate}
To identify nucleation events for bubbles decaying to $\nodim{\phi}_{\subTv} = 0$
we reflect the field around the original false vacuum $\bar{\phi} = \pi$ through the linear transform $\nodim{\phi} \to 2\pi - \nodim{\phi}$ and apply the detection algorithm to the transformed field.

We simulated ensembles of $30,000$ field evolutions for three choices of potential parameters: $\{\phi_0 = 1.35, \lambda = 1.5 \}$, $\{\phi_0 = 1.27, \lambda = 1.5 \}$, and $\{\phi_0 = 1.35, \lambda = 1.6 \}$. These combinations produce field evolutions with sufficiently rapid nucleations to have multiple events within a single simulation. The physical size of the lattice is $\phi_0^{-1} \sqrt{V_0} L = 200\sqrt{2}$ with $V_0 = 0.008 \; \phi_0^2 \mu^2$, and $N=8192$ lattice sites. The field spectrum is truncated at wavenumber index $n_{\subcut} = 256$, corresponding to a wavenumber $k_{\subcut} \approx 5.7 \phi_0^{-1} \sqrt{V_0}$. The lattice spacing is $\dd{x} = L/N$ and the discrete time step was $\dd{t} = \dd{x} / 16$.

We checked that these parameters ensure a fully resolved vacuum state. To do this, we ran ensembles of $100$ simulations initialized with a sequence of modes up to the $n_{\subcut}=256$ threshold used in our final results. First, we set $n_{\subcut}$ to the Nyquist frequency $n_{\subnyq} = N/2$, then we gradually changed $N$ such that $n_{\subnyq}$ increased each time by a factor of $2$. At each resolution, the coordinates assigned to bubble nucleation events remained constant, proving that the modes not sampled (\ie between $n_{\subcut} = 256$ and $n_{\subnyq}$) do not change the realizations significantly.

\begin{figure}[h!]
	\centering
	\includegraphics[width=1.\textwidth]{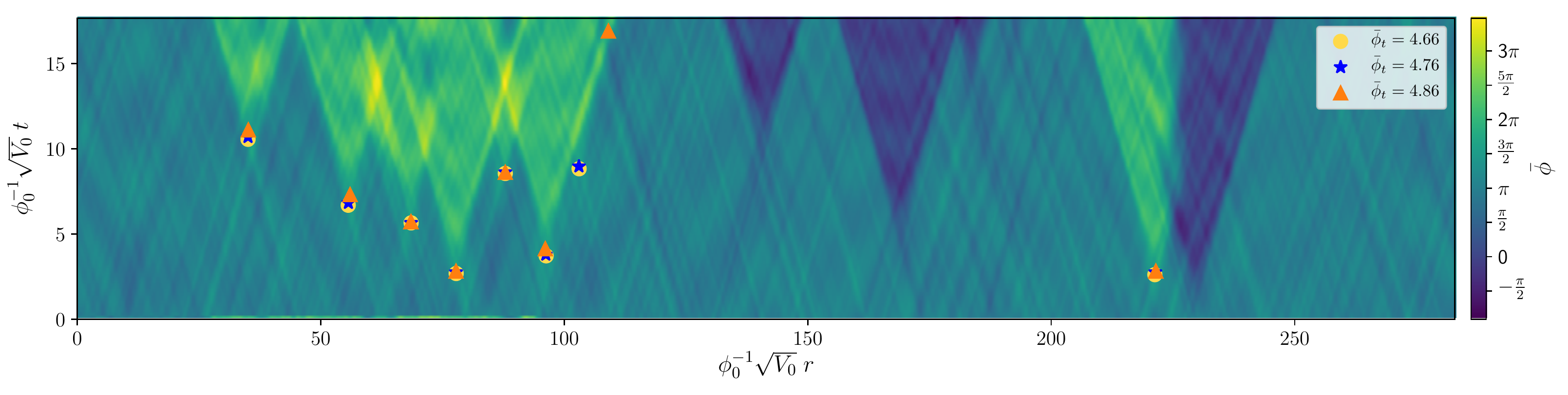}
	\caption{An example simulation with bubbles formed through both decay channels. We ran the detection algorithm to identify the nucleation sites of bubbles decaying to the true vacuum at $\nodim{\phi}_{\subTv} = 2\pi$. The detected nucleation events are shown for each threshold $\nodim{\phi}_t$. The smoothing scale used here is $R_0 = \nodim{R}/4$. We show two of the most common types of detection error. First, the orange triangle at $\phi_0^{-1}\sqrt{V_0}\;t \approx 17$ corresponding to the largest threshold is found around the edge of a bubble already accounted for, and second the background field fluctuations at $\phi_0^{-1}\sqrt{V_0}\;r \approx 110$ picked up by the two lowest thresholds.} \label{fig:egsim}
\end{figure}

To measure the efficiency of our detection algorithm, we visually inspected $100$ simulations for each combination of $\{\phi_0, \lambda\}$ and smoothing scale. We refer to a true positive event as one that has been confirmed as a true nucleation event by visual inspection; a false positive is an event that is detected but confirmed false by visual inspection. We define the efficiency as the ratio of true positive events to all positive (\ie true positive and false positive) events.  For $R_0 = \nodim{R}/4$ we estimate the efficiency for all three thresholds at $> 95\%$ when $\lambda = 1.6$ and $> 93\%$ when $\lambda = 1.5$ by counting the fraction of false positive events out of the total. Fig.~\ref{fig:egsim} shows three events which we labelled as false positives: the orange triangle at $\phi_0^{-1}\sqrt{V_0}\;t \approx 17$ is clearly within the wall of a bubble previously accounted for by the orange detector; meanwhile we considered the yellow circle and blue star at roughly $\phi_0^{-1}\sqrt{V_0}\;r \approx 110$ to correspond to background field fluctuations. Around $60\%$ of false positive events are misattributions within the physical extent of bubble walls, while the remaining $40\%$ of false positive assignments are represented by large amplitude vacuum fluctuations in the background field that eventually dissipate. Visual confirmation of results is subjective but gives a rough estimate for the error. According to our test, the total number of detected events differs by at most $4\%$ between the three thresholds, and this difference is made up of mostly false positive events. Moreover, the largest threshold $\nodim{\phi}_t$ gives up to twice as many false positives compared to the other two thresholds. The explanation for this difference is the following: since the bubble grows in amplitude with time, higher thresholds assign the events to later times than the lowest threshold and decrease the extent of the lightcone for the respective bubble, leading to erroneous detections at late times, \ie the example of the orange triangle at $\phi_0^{-1}\sqrt{V_0}\;t \approx 17$ in Fig.~\ref{fig:egsim}. We do not have an estimate for the percentage of false negatives (\ie nucleations that have not been detected). However, by visual inspection, false negative candidates appear only around nucleation sites already accounted for (e.g. two neighbouring peaks that merge during the bubble formation process can, in some cases, be detected as a single nucleation site, but sometimes as two). The number of bubble nucleation sites in such clusters is intrinsically ambiguous, and it depends on the details of how nucleation events are defined. The inclusion of such false negatives would make the correlation function computed in the next section larger at small separations. Moreover, since bubbles form with a range of center of mass velocities, their spatial coordinate also changes between different detection thresholds. Consequently, we discard as unphysical events that coincide exactly in $(t,r)$ for all three thresholds, noting that on average only about $20\%$ of these are indeed confirmed visually as nucleation sites for bubbles. We do not find a significant number of false negatives associated with bubbles that have large center of mass velocities so long as they are not associated with clusters of nucleation events, implying that relativistic length contraction does not affect the ability of our algorithm to detect bubbles.

\section{Bubble-Bubble correlation function}\label{sec:Results}

In the previous section, we made a connection between extrema in fluctuations about the false vacuum and bubble nucleation events. In analogy with the free field case, where peaks are correlated, we might expect the same to be true for bubble nucleation events. Indeed, we now show that this is the case. This is the main result of the paper.

The bubble-bubble correlation function $\xi_{bb}(t,r)$ measures the probability in excess of random that a bubble is found a fixed distance in space and time away from another. Measuring this correlation function in our data is essentially a counting problem. The bubble-bubble correlation function is defined by looking at the distribution of bubble nucleation sites. The nucleation sites are defined as the coordinate pairs $(t,r)$ where the field amplitude satisfies the conditions we enumerated in the section above. We are asking what is the probability that if a bubble is nucleated at spacetime coordinates $(0,0)$, another bubble is created within $[t, t+\Delta t) \phi_0^{-1}\sqrt{V_0} $ and $\pm [r, r+\Delta r) \phi_0^{-1}\sqrt{V_0}$. By analogy with the equal-time two-point correlator, we estimate the bubble-bubble correlation function using
\begin{equation}\label{eq:bubble_correlator}
1+\xi_{bb}(t,r) = \ev{\frac{1}{B} \sum_{i=1}^{B} \frac{B_i (t,r)}{\rho V_i (t,r))}} \, ,
\end{equation}
for a sample containing $B \in \mathbb{N}$ bubble nucleation sites. Here, $B_i(t,r)$ is the number of bubbles lying in a shell of fixed spacetime volume $V_i = 2 \Delta t \Delta r \, \phi_0^{-2}V_0$ and minimal separation $t$ and $\pm r$ in time and space, respectively, from the $i^{th}$ bubble. We have defined the binning parameters $\Delta t = 16 n_t \dd{t}$ and $\Delta r = n_r \dd{x}$ and $n_t \in \mathbb{Z} \setminus \{0\}$ and $n_r \in \mathbb{N}, n_r \leq N/2$. $\rho$ plays the role of the bubble number density for the respective sample. Averaged over all samples, we obtain our estimate for the correlation function.

We first compute the equal-time bubble correlator, given by equation \eqref{eq:bubble_correlator} where $t=0$ (defining 'equal-time' since the reference bubble is translated to $t=0$). The number density of nucleation sites $\rho$ is computed in each realization. Using an ensemble-average number density produces very similar results for the correlation function. Because the false vacuum has two decay channels, we can construct four different bubble-bubble correlation functions: $\xi_{bb}^{--}$ is the correlator between two bubbles filled with $\nodim{\phi}=0$, $\xi_{bb}^{++}$ is the correlator between two bubbles filled with $\nodim{\phi}=2\pi$, $\xi_{bb}^{+-}$ is the correlator between one bubble filled with $\nodim{\phi}=0$ and one with $\nodim{\phi}=2\pi$, and $\xi_{bb}$ is the correlator between all bubbles. Because the potential is symmetric about the false vacuum, we have $\xi_{bb}^{++} = \xi_{bb}^{--}$.  We verify that this is true in our numerical results, providing one check against biases in our algorithm for locating nucleation events. The average number density of bubbles filled with either vacuum is  equal across the ensemble, therefore $\xi_{bb}$ is the correlation function of twice more frequent events than either $\xi_{bb}^{++}$ or $\xi_{bb}^{--}$. 

\begin{figure}
	    \centering
	    \includegraphics[width=.75\textwidth]{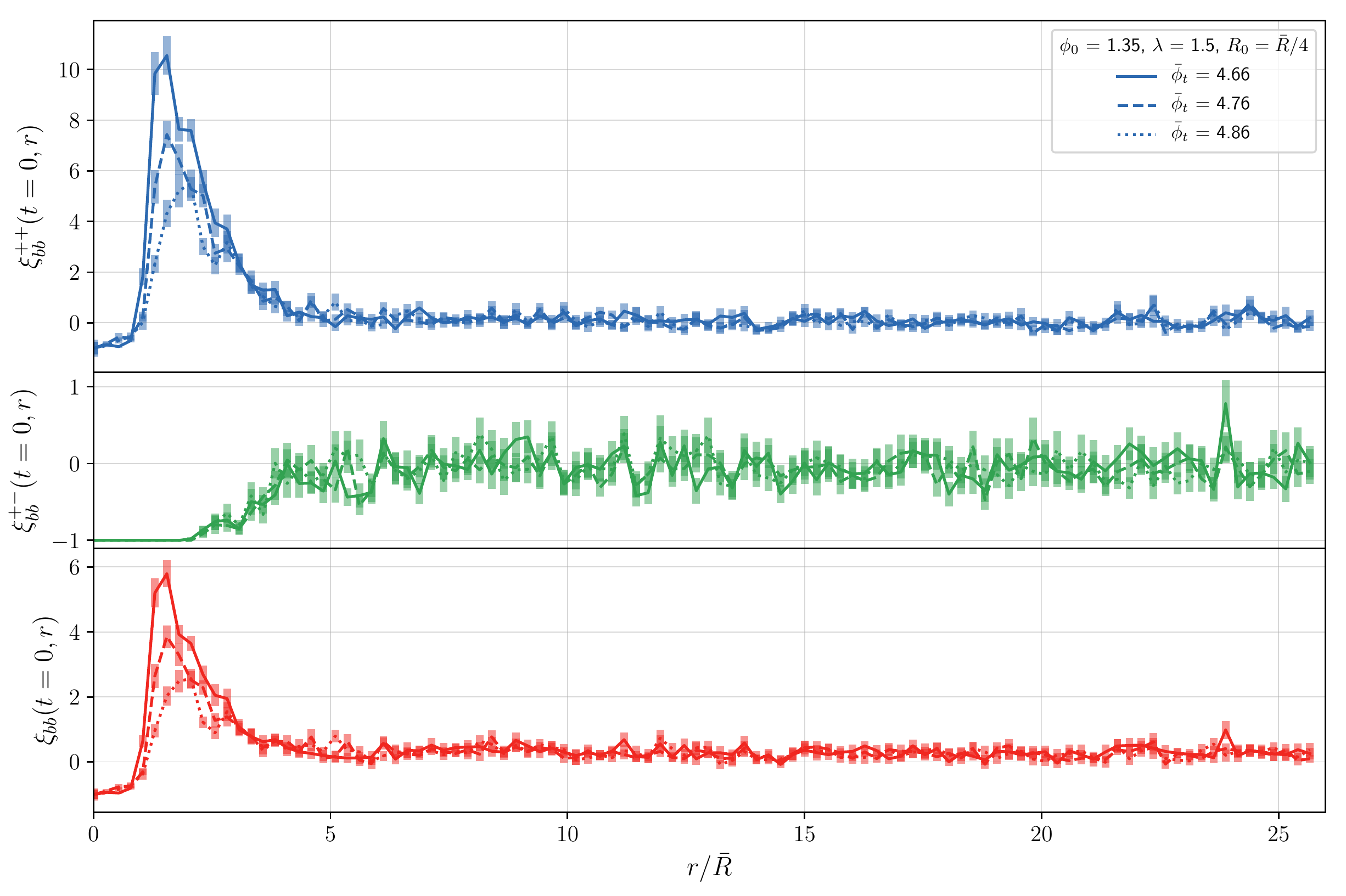}
	    \includegraphics[width=.75\textwidth]{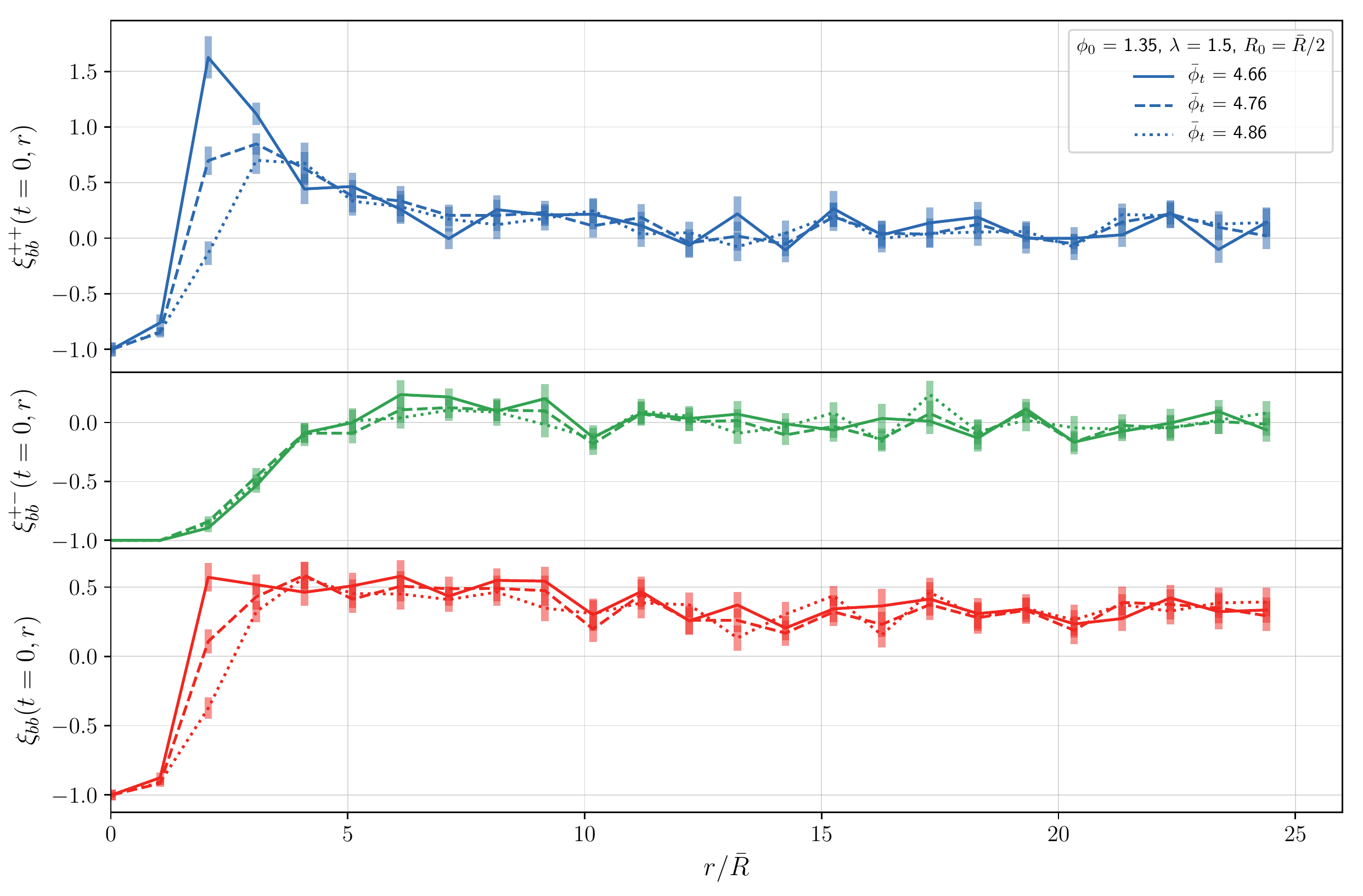}
	    \caption{In the top panel we show the three types of two-point correlation function for one choice of field parameters $\{\phi_0,\lambda\}$ and each threshold $\nodim{\phi}_t$ and filter width $R_0=\nodim{R}/4$. From top to bottom, these are the correlation function between nucleation sites of bubbles of the same kind, bubbles of opposite kinds, and the overall correlation function. In the bottom panel we show similar results for a filter width $R_0=\nodim{R}/2$. The data is binned into volumes of size $n_t = 3$ and $n_r = 10$.}\label{fig:1dCorrs_filter9}
\end{figure}

The result for the three distinct equal-time correlators is shown in Fig.~\ref{fig:1dCorrs_filter9} corresponding to the different choices of the threshold, as well as two smoothing scales, in the bubble finding algorithm. To estimate the errors, we break our ensemble of $30,000$ simulations into $30$ ensembles of $1000$ simulations each.  We compute the peak-peak correlator in each sub-ensemble, and use the RMS of the sub-ensemble means in each radial bin as an estimate of the error. Most importantly, we see from $\xi_{bb}^{++} (t, r)$ that bubbles of the same type have a statistically significant and non-trivial positive correlation. Nucleation events are significantly correlated, and therefore cluster, over a distance of order a few times the initial size of bubbles when they nucleate, $\nodim{R}$. Bubbles of opposite type have a negative correlation $\xi_{bb}^{+-} (t, r)$, which follows from the fact that finding a large peak and a large trough near each other in the field is a rare event. The dependence of $\xi_{bb}^{++} (t, r)$ on the threshold chosen in the bubble finding algorithm is significant, but the qualitative features of the correlation function remain unaltered. The mismatch is a result of the different definitions for a nucleation site that each threshold implies. Some signals in the field are detected as bubbles for one choice of $\nodim{\phi}_t$, but not for another. Moreover, different thresholds assign a different location for each bubble nucleation event (\eg a larger choice of threshold implies a relatively later stage in the process of formation for a bubble).  Therefore events corresponding to the same bubble might show up in different bins of $\xi_{bb}$, depending on $\nodim{\phi}_t$. Note, however, that the correlation function peaks at roughly the same scale regardless of the choice of threshold. We also compare two choices of smoothing scale, equal to $\nodim{R}/4$ (top panel of Fig.~\ref{fig:1dCorrs_filter9}) and $\nodim{R}/2$ (bottom panel of Fig.~\ref{fig:1dCorrs_filter9}). As expected, for a larger smoothing scale, we loose the ability to resolve correlations at short distances. Smoothing is equivalent to coarse-graining the lattice, merging clusters into single, stand-alone bubbles. Nevertheless, there is still a significant correlation for identical bubbles, and a significant anti-correlation for non-identical bubbles.

\begin{figure}
	\centering
	\includegraphics[width=0.75\textwidth]{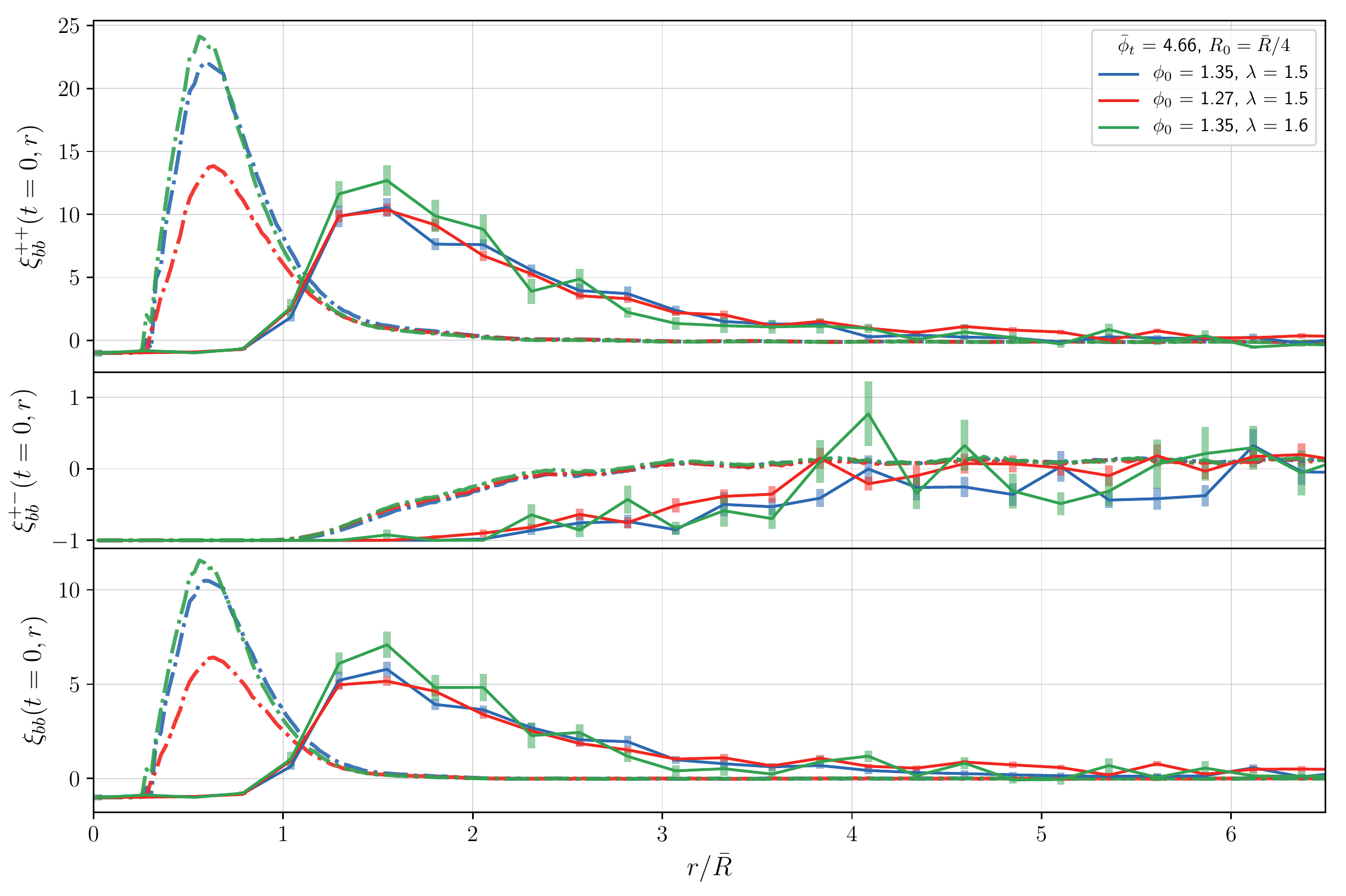}
	\caption{The bubble-bubble correlator $\xi^{++}_{bb}(t,r)$ (solid curves, with error bars) versus the peak-peak correlator $\xi_{pk}(r)$ (dotted curves, no error bars) for all three combinations of parameters $\{\phi_0, \lambda\}$ and fixed threshold $\nodim{\phi}_t$ and smoothing scale $R_0 = \nodim{R}/4$. The qualitative agreement between the curves is apparent except that the peak-peak correlator picks up amplitude at $r=R_0$, and the bubble-bubble correlator at $r=\nodim{R}$, which is expected. The mixed two-point functions $\xi_{bb}^{+-}$ and $\xi_{pk}^{+-}$ show anti-correlation over scales up to the correlation length of their counterparts $\xi_{bb}^{++}$ and $\xi_{pk}^{++}$. While the $\xi_{pk}$ is sensitive to the size of the fluctuations at initialization $\phi_0$, the $\xi_{bb}$ only reacts noticeably to the $\lambda$ parameter. Data is binned into boxes of size $n_t = 3$ and $n_r = 10$.}\label{fig:1dCorrs_BBvsPK}
\end{figure}

In Fig.~\ref{fig:1dCorrs_BBvsPK}, we compare the bubble-bubble correlation function to the peak-peak correlation function obtained for a free massive scalar with mass set by the curvature about the false vacuum. We choose the lowest threshold imposed in the bubble finding algorithm, and compute the correlation function for three choices of potential parameters. The qualitative similarity between the bubble-bubble and peak-peak correlator is apparent (with an exclusion region, peak, and decay), and for this choice of threshold there is even reasonably good agreement in the amplitude. 

There are, however, a few important differences. Increasing the threshold for the free field reduces the number density of peaks, thus increasing the amplitude of the correlation function. For a perfect bubble detection algorithm, there should be a critical value for the threshold beyond which the bubble finding algorithm will be insensitive to the threshold choice, at least until it approaches the true vacuum.  This is because once the threshold is high enough for bubbles to form, increasing the threshold should only displace the nucleation event in space and time. However, as commented on above, the highest threshold has the largest false-positive rate. This systematic has the effect of suppressing the correlation function with increasing threshold. This is the reason we have chosen the lowest threshold, which has the lowest false-positive rate, and should more faithfully represent the true bubble-bubble correlator. Another difference is the change in amplitude of the correlation function for different potential choices.  For peaks in the free field, larger values of $\phi_0$ decrease the amplitude of vacuum fluctuations relative to the width of the false vacuum minimum, decreasing the number density of peaks exceeding our threshold and leading to stronger clustering (comparing the red and blue curves in Fig.~\ref{fig:1dCorrs_BBvsPK}). For bubbles, the nucleation rate does decrease with increasing $\phi_0$ as expected. However, the clustering of bubbles for the two different values of $\phi_0$ we probed is unchanged,
 as we see when comparing the correlation functions for potentials with $\lambda = 1.5$ and $\phi_0 = 1.35$ vs $\phi_0 = 1.27$ in Fig.~\ref{fig:1dCorrs_BBvsPK}. The fractional increase in the peak correlation function is far larger than for the bubble correlation function. On the other hand, increasing $\lambda$ at fixed $\phi_0$ makes both peaks and bubbles more rare on the lattice (in fixed length units of $\phi_0^{-1}\sqrt{V_0}$), manifesting as a higher amplitude correlation function in both cases. Finally, the peak-peak correlator has a maximum closer to the filter width than the bubble-bubble correlator. This is due to the fact that bubbles are somewhat larger than the filter, and cannot cluster on scales smaller than their size (which is roughly 4 times the filter width).

\begin{figure}
    \centering
    \includegraphics[width=\textwidth]{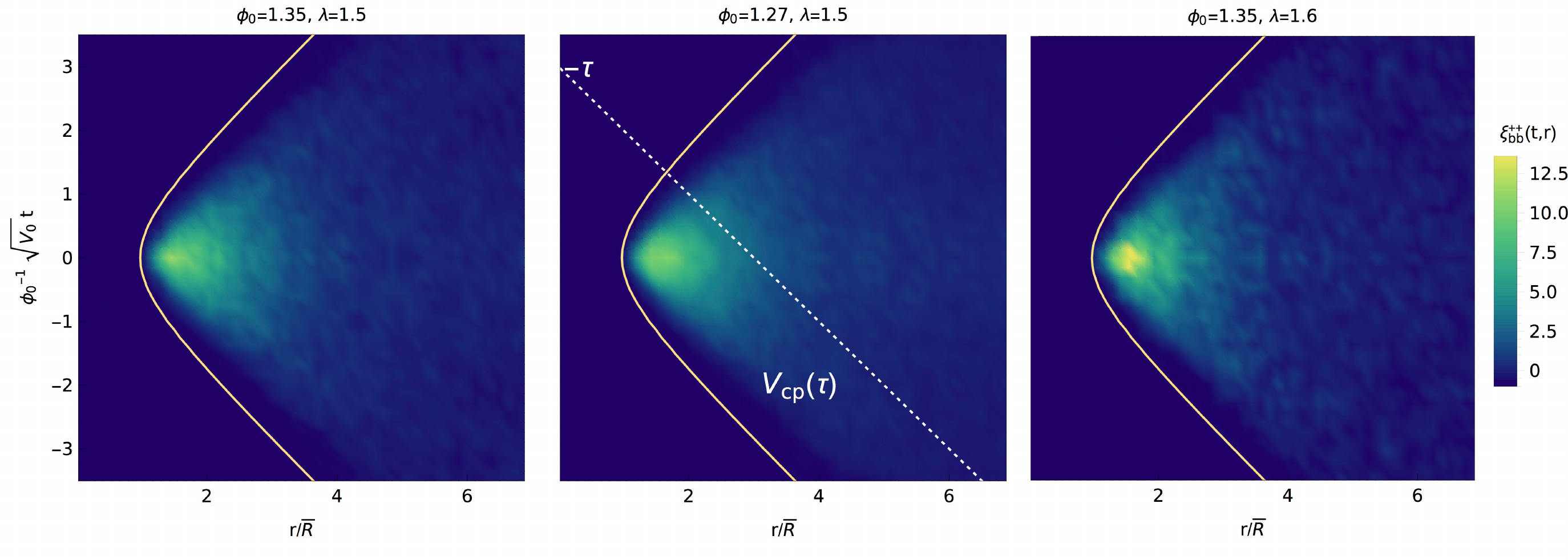}
    \caption{The $2$-dimensional bubble-bubble correlator $\xi^{++}_{bb}(t,r)$ for all three combinations of parameters $\{\phi_0, \lambda\}$ and smoothing scale $R_0=\nodim{R}/4$. The correlation function implies clustering in both $r$ and $t$ directions. Data is binned into $n_t = 3$ and $n_r = 5$ intervals. Errors are not shown and symmetry is assumed for negative $r$. In each panel, we overplot the trajectory of the (time-symmetric) average bubble wall; a hyperbola with radius $\nodim{R}$. In the centre panel, we depict the past lightcone of a hypothetical observer at the origin of coordinates. This observer will have causal access to bubbles that nucleate in the spacetime volume $V_{cp} (\tau)$ between the bubble wall and the past lightcone.}\label{fig:2d_corr}
\end{figure}

In Fig.~\ref{fig:2d_corr} we show the ensemble-averaged $\xi_{bb}^{++}$ correlator in both space and time (Eq.~\eqref{eq:bubble_correlator}) for the three parameter choices for $\{\phi_0, \lambda\}$. The contours describe the increased (or decreased) probability for a bubble to nucleate at some point $\{ t,r \}$ given a bubble whose nucleation centre is at $\{t=0, r=0\}$ In the standard picture of vacuum decay, the bubble at the origin would nucleate at $t=0$ with a size of $r = \nodim{R}$ and the wall would expand on the hyperbolic trajectory shown in Fig.~\ref{fig:2d_corr}. We also show the past-directed hyperboloid for reference. Note that the correlation function has no structure inside of the bubble wall hyperboloid, which is a good check on the bubble detection algorithm: nucleation events cannot happen inside of a bubble or in its causal past. Additionally, note that the correlator is time-symmetric, but it does not have a symmetry with respect to Lorentz boosts. Although the vacuum state is Lorentz invariant, a configuration with a bubble nucleation event is not. From the perspective of the peak-peak correlator in a Gaussian random field, we could understand the decay of the correlation function in time as being due to the finite lifetime of peaks. Therefore, clustering should occur in a concentrated region of both space and time. We now discuss the phenomenological consequences of a space-time bubble correlation function.
 
\section{Phenomenological implications}\label{sec:discussion}

\begin{figure}
	\centering
	\includegraphics[width=.5\textwidth]{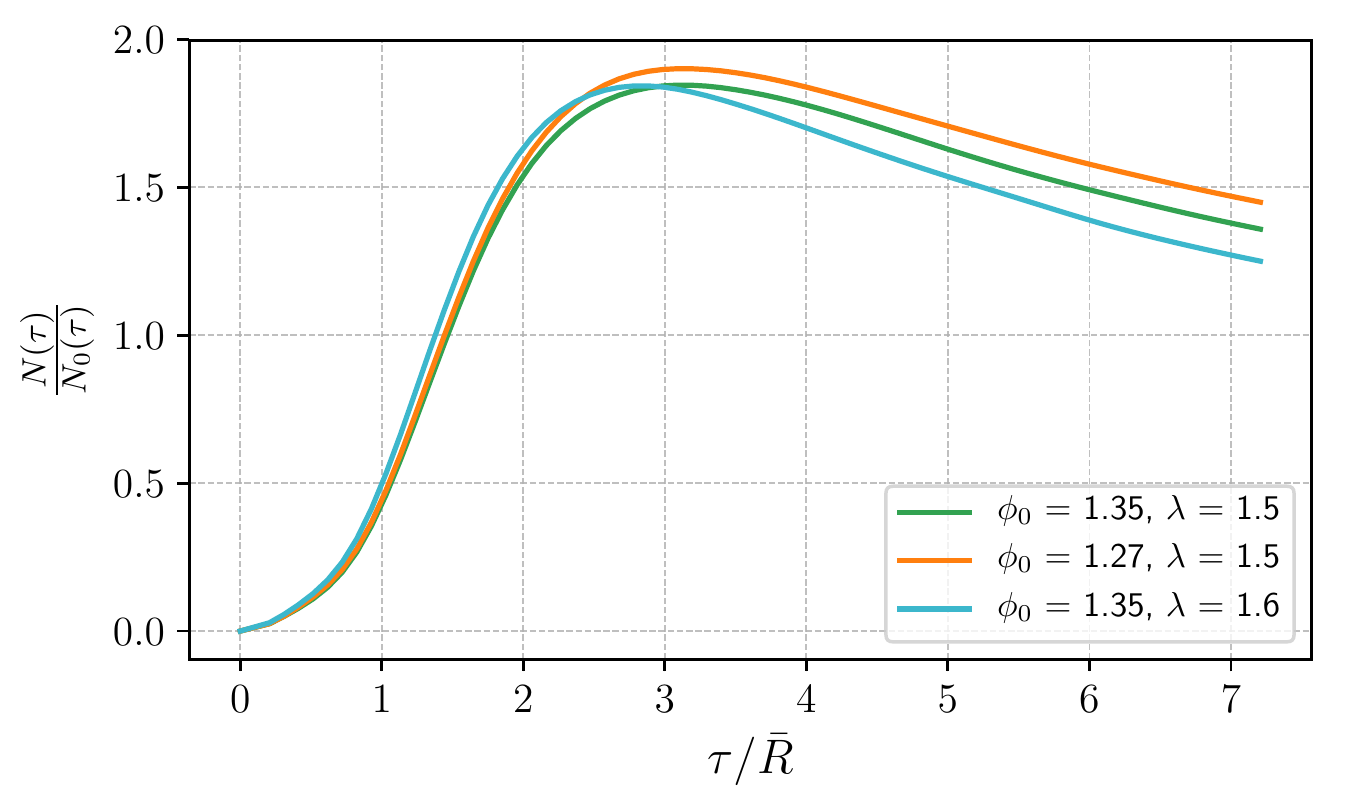}
	\caption{The fractional change in the number of observed collisions to the causal past of an observer situated at time $\tau$ at the centre of the reference bubble for the correlation functions in Fig.~\ref{fig:2d_corr}. The distribution peaks once the peak of the correlation function $\xi_{bb}^{++}$ is entirely to the observer's past.}\label{fig:fractional_change_in_N}
\end{figure}

In this section we briefly highlight the phenomenological implications of a non-trivial bubble-bubble correlation function. We can begin to understand these implications by considering Fig.~\ref{fig:2d_corr}. Given a bubble at some location in spacetime, there is an enhanced probability of another bubble nucleating nearby in space and time. This is in contrast to the typical assumption that bubble nucleation events occur with equal probability in any region of spacetime. Consider an observer at the origin of the central bubble in Fig.~\ref{fig:2d_corr}, which originates at $\{r=0, t=0\}$. If another bubble were to nucleate from the false vacuum region outside of the central bubble, and in the causal past of the observer at proper time $\tau$ (the spacetime volume in Fig.~\ref{fig:2d_corr} between the hyperbola and the past lightcone), then they would have causal access to a bubble collision. The average number of collisions $N(\tau)$ is an integral of the nucleation rate $\ev{\rho}$ over the spacetime volume available to nucleate colliding bubbles. The existence of a non-trivial bubble correlation function can lead to an enhancement or suppression in the average number of collisions given by:
\begin{equation}\label{eq:change_numbubs}
   \frac{N(\tau)}{N_0(\tau)} = \frac{\int_{V_{cp}(\tau)} \dd{t} \dd{r} \ev{\rho} (1+\xi(r,t))  }{\int_{V_{cp}(\tau)} \dd{t} \dd{r} \ev{\rho} } = \langle(1+\xi(r,t))\rangle_{V_{cp}(\tau)},
\end{equation}
where $V_{cp}(\tau)$ denotes the spacetime volume that is both in the causal past of an observer at proper time $\tau$ and also outside of their bubble. In the second equality we have assumed that $\ev{\rho}$ is independent of space and time, in which case the enhancement or suppression in the number of observed collisions is simply the average of the correlation function over $V_{cp}(\tau)$. From the definition of the correlation function, if an observer had access to the entire volume of space and time then this volume average would be unity. Crucially, for an observer at fixed proper time, the result of evaluating \eqref{eq:change_numbubs} will be non-trivial. For the examples shown in Fig.~\ref{fig:2d_corr}, we expect an enhancement in the number of observed collisions that peaks at a time of order $\tau \sim \nodim{R}$ (when the peak in the correlator is encompassed by $V_{cp}(\tau)$) and subsequently decays with $\tau$ as the volume average dilutes the region of significant correlation. We show this in Fig.~\ref{fig:fractional_change_in_N} by numerically integrating the amplitude of $\xi_{bb}^{++}$ for each of the cases shown in Fig.~\ref{fig:2d_corr}. The data used is the raw binned data for the correlation function smoothed with a 1+1-dimensional Gaussian filter of size $3\times 16\dd{t}\phi_0^{-1} \sqrt{V_0} \approx 0.105$ and $5\times \dd{x}\phi_0^{-1} \sqrt{V_0} \approx 0.175$ in the time and space dimensions, respectively.

There are two contexts where bubble-bubble correlations could have interesting phenomenological consequences. The first is percolating phase transitions that produce observable gravitational waves~\cite{PhysRevD.47.4372,PhysRevLett.69.2026,Kamionkowski1994}. The gravitational wave spectrum peaks at a frequency set by the average size of bubbles at the time when percolation occurs~\cite{PhysRevD.47.4372,PhysRevLett.69.2026,Kamionkowski1994,Leitao_2016}. In previous literature, it was assumed that bubbles nucleate at random positions, in which case the average size of bubbles at the time of collision is simply the duration of the phase transition. In the presence of a bubble-bubble correlation function such as the one in Fig.~\ref{fig:2d_corr}, bubbles are likely to form significantly closer together than in the standard picture. Understanding the consequences in detail is beyond the scope of the present work, but we can speculate that there will be a secondary peak in the gravitational wave spectrum corresponding to the scale over which bubbles are correlated, namely their initial radius. There may also be implications for the angular power spectrum of the stochastic gravitational wave background~\cite{Geller:2018mwu}. The second scenario where bubble-bubble correlations could be relevant is bubble collisions produced during eternal inflation~\cite{PhysRevD.76.123512,Aguirre2007a,Chang:2007eq}. In this case, the enhancement in the expected number of collisions due to the bubble-bubble correlator described above will enhance the probability of observing the signatures of collisions in the CMB~\cite{PhysRevLett.107.071301}. Again, we defer a detailed investigation of the consequences of such an enhancement to future work.

\section{Conclusion}\label{sec:conclusion}

The semi-classical stochastic treatment of vacuum decay makes it possible to investigate issues related to the dynamics of bubble formation that have previously been inaccessible. In this work, we showed that, analogously to the process of biased galaxy formation, the nature of the fluctuations around the false vacuum state gives rise to clusters of bubble nucleation sites. To our knowledge this is the first time biasing has been investigated in the context of bubble formation in vacuum decay. We found that a significant correlation exists between the nucleation sites of thick-wall bubbles in a single scalar field theory. A finite correlation length between bubble nucleation sites implies a greater chance of collisions between bubbles nearby in space and time. We briefly speculated that this could lead to new features in the stochastic gravitational wave spectrum associated with first-order phase transitions in the early Universe, and could increase the probability of observing the collisions between bubbles in the scenario of eternal inflation. 

There are several directions for future work. Extending the present work to more spatial dimensions and confirming that the bubble-bubble correlation function is qualitatively similar to that presented here is clearly an important first step. A detailed investigation of the impact of the bubble-bubble correlation function on the phenomenology of first order phase transitions in the early Universe should be undertaken. This could lead to new targets for future gravitational wave observatories such as LISA. In another direction, the potential we studied is motivated by the program of simulating vacuum decay in cold atom systems. These systems should exhibit a bubble-bubble correlation function analogous to the one investigated here, and could constitute an important observable. Exploring how the bubble-bubble correlation function can be used to determine the vacuum statistics could also be useful in understanding the results of these experiments. More broadly, the bubble-bubble correlation function is just one example of how the real-time simulations of vacuum decay might be used. Future work will explore a number of other applications. 

\acknowledgements
We thank A. Jenkins, H. Peiris, A. Pontzen, and S. Weinfurtner for helpful comments and discussion. MCJ is supported by the National Science and Engineering Research Council through a Discovery grant. This research was supported in part by Perimeter Institute for Theoretical Physics. Research at Perimeter Institute is supported by the Government of Canada through the Department of Innovation, Science and Economic Development Canada and by the Province of Ontario through the Ministry of Research, Innovation and Science. 

\bibstyle{JHEP}
\bibliography{references.bib}

	\end{document}